\documentclass[acmlarge]{acmart}

\usepackage[font=small,labelfont=bf]{caption}
\usepackage{graphicx, balance}
\usepackage{caption,subcaption}
\usepackage{linguex}
\usepackage[linguistics,edges]{forest}
\usepackage{enumitem}
\usepackage{booktabs}
\usepackage{multirow, makecell}
\usepackage{algorithmic}
\usepackage{url}
\usepackage{csquotes} 
\usepackage{arydshln} 
\usepackage{color, xcolor, soul}
\usepackage{makecell}
\usepackage{cite}
\usepackage{etoolbox}
\usepackage{threeparttable}
\usepackage{hyperref}
\usepackage{pifont}

\usepackage{rotating}
\usepackage{array}
\usepackage{booktabs}
\usepackage{multirow}
\usepackage{tabularx}

\usepackage{tikz}
\usetikzlibrary{shapes.geometric, arrows}

\usepackage[most]{tcolorbox}

\tcbset{textmarker/.style={%
        enhanced,
        parbox=false,
        boxrule=0mm,
        boxsep=0.5mm,
        arc=0mm,
        outer arc=0mm,
        left=1mm,
        right=0mm,
        top=0mm,
        bottom=0mm,
        toptitle=1mm,
        bottomtitle=1mm,
        oversize,
        frame hidden}}
\newtcolorbox{promptBox}{textmarker,
    borderline west={1pt}{0pt}{black},
    colback=black!5!white}
\newtcolorbox{takeawaysBox}{textmarker,
    borderline west={1pt}{0pt}{blue},
    colback=black!0!white}
\newtcolorbox{limitationBox}{textmarker,
    borderline west={1pt}{0pt}{red},
    colback=black!0!white}

\newcommand{\takeaways}[1]{\begin{takeawaysBox} \textbf{Takeaways.} #1 \end{takeawaysBox}}
\newcommand{\limitations}[1]{\begin{limitationBox} \textbf{Limitations.} #1 \end{limitationBox}}

\usepackage{fancyhdr}
\makeatletter
\def\@authorsaddresses{}
\makeatother

\newcolumntype{C}[1]{>{\centering\arraybackslash}p{#1}}

\newcommand{\deleted}[1]{}

\usepackage{lscape}
\begin{document}

\title{From ML to LLM: Evaluating the Robustness of Phishing Webpage Detection Models against Adversarial Attacks}
\author{Aditya Kulkarni}
\email{aditya.kulkarni@iitdh.ac.in}
\affiliation{%
 \institution{Indian Institute of Technology (IIT) Dharwad}
 \country{India}}

\author{Vivek Balachandran}
\email{vivek.b@singaporetech.edu.sg}
\affiliation{%
 \institution{Singapore Institute of Technology}
 \country{Singapore}}

\author{Dinil Mon Divakaran}
\email{dinil_divakaran@i2r.a-star.edu.sg}
\affiliation{%
 \institution{Institute for Infocomm Research (I$^2$R), A*STAR}
 \country{Singapore}}

\author{Tamal Das}
\email{tamal@iitdh.ac.in}
\affiliation{%
 \institution{Indian Institute of Technology (IIT) Dharwad}
 \country{India}}

\definecolor{StartStopColor}{HTML}{FFFFFF} 
\definecolor{InputOutputColor}{HTML}{FFFFFF} 
\definecolor{ProcessColor}{HTML}{FFFFFF} 
\definecolor{DecisionColor}{HTML}{FFFFFF} 
\definecolor{textcolor}{HTML}{000000} 
\definecolor{processtextcolor}{HTML}{FFFFFF} 

\tikzstyle{startstop} = [rectangle, rounded corners, minimum width=1.75cm, minimum height=1cm, text centered, draw=black, text=textcolor, fill=StartStopColor]
\tikzstyle{io} = [trapezium, trapezium left angle=70, trapezium right angle=110, text width=4.5cm, minimum height=1cm, text centered, draw=black, fill=InputOutputColor, text=textcolor]
\tikzstyle{process} = [rectangle, text width=5.5cm, minimum height=1cm, text centered, draw=black, fill=ProcessColor, text=textcolor]
\tikzstyle{decision} = [diamond, minimum width=3cm, minimum height=1cm, text centered, draw=black, fill=DecisionColor]
\tikzstyle{arrow} = [thick,->,>=stealth]

\tikzstyle{io_1} = [trapezium, trapezium left angle=70, trapezium right angle=110, text width=6cm, minimum height=1cm, text centered, draw=black, fill=InputOutputColor, text=textcolor]
\tikzstyle{process_1} = [rectangle, text width=5cm, minimum height=1cm, text centered, draw=black, fill=ProcessColor, text=textcolor]

\begin{abstract}
  Phishing attacks attempt to deceive users into stealing sensitive information, posing a significant cybersecurity threat. Advances in machine learning (ML) and deep learning (DL) have led to the development of numerous phishing webpage detection solutions, but these models remain vulnerable to adversarial attacks. Evaluating their robustness against adversarial phishing webpages is essential. Existing tools contain datasets of pre-designed phishing webpages for a limited number of brands, and lack diversity in phishing features.
  
  To address these challenges, we develop \texttt{PhishOracle}, a tool that generates adversarial phishing webpages by embedding diverse phishing features into legitimate webpages. We evaluate the robustness of three existing task-specific models---Stack model, VisualPhishNet, and Phishpedia---against \texttt{PhishOracle}-generated adversarial phishing webpages and observe a significant drop in their detection rates. In contrast, a multimodal large language model (MLLM)-based phishing detector demonstrates stronger robustness against these adversarial attacks but still is prone to evasion. Our findings highlight the vulnerability of phishing detection models to adversarial attacks, emphasizing the need for more robust detection approaches. Furthermore, we conduct a user study to evaluate whether \texttt{PhishOracle}-generated adversarial phishing webpages can deceive users. The results show that many of these phishing webpages evade not only existing detection models but also users.
\end{abstract}



\maketitle

\section{Introduction}
\label{sec:Introduction}
Phishing attacks extract sensitive information from targeted users. Attackers create phishing webpages and share URLs with users via social media platforms, emails, etc. These requests deceive users into clicking on the URLs and redirect them to phishing webpages where they submit sensitive information. Attackers use this sensitive information to launch identity theft attacks or use information to exploit opportunities for financial profits. In Q1 $2024$, the APWG~\citep{APWG_REPORT_Q1_2024} recorded 963,994 phishing attacks, with the social media platforms being the most targeted sector, accounting for $37.4 \%$ of the attacks. Recent years have witnessed improvements in phishing webpage detection approaches with the use of machine learning (ML) and deep learning (DL) methods.

The ML-based phishing webpage detection solutions~\citep{shirazi2018kn0w, jain2018phish, niakanlahiji2018phishmon, rao2020catchphish} consist of several stages. Initially, datasets containing both phishing and legitimate samples, including URLs, HTML content, and screenshots, are collected from publicly available repositories. Subsequently, relevant features are extracted based on the URL (such as URL length, presence of \texttt{@}, HTTPS, hyphens in the URL, etc.), webpage content (such as internal and external hyperlinks, CSS styles, pop-up logins, \texttt{<form action="">} fields, etc.), and third-party features (such as PageRank, Google Index, etc.). Models such as Random Forest (RF), Decision Tree (DT), Support Vector Machine (SVM), and many others are trained on these datasets, with performance evaluated using metrics like accuracy, precision, recall, and F1-Score. Finally, the model's efficiency is validated on newly unseen phishing webpages.

With the continual progress in DL techniques, there is an enhanced capability to analyze webpage screenshots and logos by employing state-of-the-art computer vision (CV) models~\citep{fu2006detecting, afroz2011phishzoo, abdelnabi2020visualphishnet, lin2021phishpedia, liu2022inferring}. DL techniques are utilized to identify brand logos on webpage screenshots, comparing them with a reference brand list. This comparison considers similarity scores and domain analysis to classify the webpage as phishing or legitimate.

Recently, large language models (LLMs) have gained significant attention and adoption across various industries. These models pretrained on massive datasets are capable to understand and generate text, code, images, and videos. Therefore, the past two years have seen increasing application of LLMs to address different challenges in cybersecurity~\citep{divakaran2024large}. In the context of phishing, a recent work by Lee~\textit{et al.}~\citep{lee2024multimodal} proposes an LLM-based pipeline that identifies the targeted brand by analyzing both HTML contents and webpage screenshots. The results show that multimodal large language models (MLLMs) are helpful in outperforming a state-of-the-art solution, namely VisualPhishNet~\citep{abdelnabi2020visualphishnet}, which uses CV for  similarity-based phishing detection.

Analyzing existing ML and DL-based phishing webpage detection approaches reveals significant challenges. Initially, researchers face time-consuming efforts for dataset collection containing phishing webpages with diverse phishing features. Furthermore, the short-lived nature of phishing webpages necessitates the extraction of both content-based and third-party-based features. Subsequently, it is essential to evaluate the robustness of these detection models to ensure they can accurately classify new phishing webpages. Recent works~\citep{apruzzese2022mitigating, apruzzese2023real, lee2023attacking, ji2024evaluating, charmet2024vortex} emphasize the importance of assessing the models against adversarial phishing webpages to determine their performance in detecting such sophisticated threats.

Existing phishing tools (BlackEye~\citep{BlackEye}, ZPhisher~\citep{ZPhisher}, and ShellPhish~\citep{ShellPhish}) contain pre-designed phishing webpages datasets that can be used to evaluate the robustness of phishing webpage detection models. However, these datasets are limited to a small number of brands and lack diversity in phishing features. Moreover, these tools cannot generate multiple versions of phishing webpages with diverse sets of phishing features. We tested several brands, such as Yahoo, Amazon, Instagram, Facebook, and Netflix from BlackEye~\citep{BlackEye}. However, since the tool provides repeated URLs with each execution, many of these URLs were already blocked, rendering the adversarial pages ineffective for assessing evasion attacks.

To overcome these limitations, we develop an automated tool---\texttt{PhishOracle}---that \textit{generates} adversarial phishing webpages by embedding diverse phishing features into legitimate webpages. \texttt{PhishOracle} parses webpage content and randomly selects a set of $12$ content-based and $5$ visual-based phishing features (refer Table~\ref{tab:Phishing_Features}), enabling the creation of diverse phishing webpages without being limited to a predefined set of brands. Due to this randomized process, \texttt{PhishOracle} can generate multiple phishing webpages for a single legitimate webpage, each with a unique set of phishing features. Furthermore, \texttt{PhishOracle} is extensible, allowing the integration of new phishing features to adapt to evolving phishing threats. These generated adversarial phishing webpages are used to evaluate the robustness of existing phishing webpage detection models.

In this paper, we evaluate the robustness of phishing detection systems---that use conventional ML (Stack model~\citep{li2019stacking}), DL (VisualPhishNet~\citep{abdelnabi2020visualphishnet}, Phishpedia~\citep{lin2021phishpedia}), and LLM-based phishing detector~\citep{lee2024multimodal}---against \texttt{PhishOracle}-generated adversarial phishing webpages. Our results reveal that traditional phishing detection models experience a significant drop in detection rate against the adversarial phishing webpages. While LLM-based phishing detector demonstrates higher resilience compared to these models, but it still experiences a decline in detecting adversarial phishing pages. We summarize our contributions:

\begin{enumerate}
    \item We propose \texttt{PhishOracle}, a phishing webpage \textit{generator} capable of producing adversarial phishing webpages by randomly embedding content-based and visual-based phishing features into legitimate webpages. We carry out comprehensive evaluations to evaluate the robustness of ML, DL and LLM-based phishing webpage detectors using \texttt{PhishOracle}-generated pages. To the best of our knowledge, this is the first work to do so with one tool.
    \item We conduct a user study to evaluate the effectiveness of \texttt{PhishOracle}-generated phishing webpages generated  to deceive users. The experiment results demonstrate that on average $\sim$48 $\%$ of these generated phishing webpages are incorrectly classified as legitimate by the users.
    \item We further validate the effectiveness of adversarial phishing webpages generated by \texttt{PhishOracle} by testing them against $90+$ security vendors on VirusTotal.
    \item Finally, we contribute a dataset containing 9,067 legitimate webpage screenshots, on which we manually labelled the identity logos. The \texttt{PhishOracle} code base, web app, generated phishing webpages, and survey screenshots are available on our GitHub repositories~\citep{PHISHORACLE_GITHUB, PHISHORACLE_WEBAPP_GITHUB}.
\end{enumerate}

Additionally, we develop \texttt{PhishOracle} web app, which allows users to selectively incorporate both content-based and visual-based phishing features into a legitimate webpage. Compared to the existing phishing webpage dataset tools, \texttt{PhishOracle} offers a broader range of features, including content-based manipulations such as pop-up logins and logo transformation techniques like adjusting opacity, adding watermarks, and more--capabilities that existing tools lack. Table~\ref{tab:Tools_Comparison} provides a comparative analysis of phishing webpages dataset tools and \texttt{PhishOracle}, highlighting the range of phishing features each tool supports. \texttt{PhishOracle} is a \textit{grey-hat} tool, primarily designed to validate existing phishing webpage detection solutions, which are inherently white-hat in nature. However, its capability to generate phishing webpages introduces a grey-hat aspect to its functionality.

The rest of this paper is structured as follows: Section~\ref{sec:Related_Work} reviews existing phishing webpage dataset toolkits drawing a comparison with \texttt{PhishOracle} and discusses ML, DL and LLM models for phishing webpage detection. Section~\ref{sec:Threat_Model} defines the threat model, outlining the adversarial capabilities and objectives considered in our evaluation. Section~\ref{sec:Proposed_Phishing_Webpage_Generation_Tool_PhishOracle} discusses \texttt{PhishOracle}, a tool that generates adversarial phishing webpages by adding randomly selected content-based and visual-based phishing features into legitimate webpages. Section~\ref{sec:Experimental_Setup} details the performance metrics and datasets used, including the creation of evasion dataset with \texttt{PhishOracle}. It also evaluates robustness of the Stack model, VisualPhishNet, Phishpedia, as well as the LLM-based phishing detector. Moreover, the section presents the evaluation of tools from different security vendors against these adversarial phishing webpages. Section~\ref{sec:User_Study_and_Contributions} describes a user study assessing the effectiveness of \texttt{PhishOracle}-generated adversarial phishing webpages in deceiving users based on the visual appearance. It also outlines \texttt{PhishOracle} web app, which is designed to generate phishing webpages. Section~\ref{sec:Limitations} discusses the limitations of our approach. Finally, Section~\ref{sec:Conclusion_and_Future_Scope} concludes the paper by discussing the performance of phishing webpage detection models on \texttt{PhishOracle}-generated adversarial phishing webpages.

\begin{table*}[th]
    \caption{Existing Phishing Tools vs \texttt{PhishOracle}}
    \label{tab:Tools_Comparison}
    \begin{tabular}{lcccccccccccccccc}
    \toprule
        \multirow{15}{*}{\textbf{Tools}} & \multicolumn{15}{c}{\textbf{Phishing Features}} & \multirow{15}{*}{\makecell{\textbf{Static (S)/} \\ \textbf{Dynamic (D)}}} \\ \cmidrule{2-16}
        ~ & \rotatebox{90}{\texttt{action="local.php"}} & \rotatebox{90}{Save or Mail Credentials}  & \rotatebox{90}{Added/Removed \texttt{<script>} tags } & \rotatebox{90}{Local CSS} & \rotatebox{90}{Local JS} & \rotatebox{90}{Modified Hypertext References} & \rotatebox{90}{Font Style} & \rotatebox{90}{Disable \texttt{<a>} tags} & \rotatebox{90}{Disable other login \texttt{<button>}} & \rotatebox{90}{Popup Login} & \rotatebox{90}{Disabled Right Click} & \rotatebox{90}{Disabled CTRL and Fn} & \rotatebox{90}{Body Opacity} & \rotatebox{90}{Replace blank space with character } & \rotatebox{90}{Logo Transformations} & \\ \midrule
        BlackEye~\citep{BlackEye} & \ding{51} & \ding{51} & \ding{51} & ~ & ~ & ~ & \ding{51} & ~ & ~ & ~ & ~ & ~ & ~ & ~ & ~ & \textbf{S} \\
        ZPhisher~\citep{ZPhisher} & \ding{51} & \ding{51} & \ding{51} & \ding{51} & \ding{51} & ~ & ~ & ~ & ~ & ~ & ~ & ~ & ~ & ~ & ~ & \textbf{S} \\ 
        ShellPhish~\citep{ShellPhish} & \ding{51} & \ding{51} & \ding{51} & \ding{51} & \ding{51} & \ding{51} & \ding{51} & ~ & ~ & ~ & ~ & ~ & ~ & ~ & ~ & \textbf{S} \\ 
        \textbf{\texttt{PhishOracle}} & \ding{51} & \ding{51} & \ding{51} & \ding{51} & \ding{51} & \ding{51} & \ding{51} & \ding{51} & \ding{51} & \ding{51} & \ding{51} & \ding{51} & \ding{51} & \ding{51} & \ding{51} & \textbf{D} \\ \bottomrule
    \end{tabular}
\end{table*}

\section{Related Work}
\label{sec:Related_Work}
In this section, we summarize the existing datasets of pre-designed phishing webpages, which are limited to a small set of brands. We contrast these with \texttt{PhishOracle}, which \textit{generates} phishing webpages by embedding diverse phishing features into legitimate webpages. Additionally, we review existing ML, DL and LLM-based phishing webpage detection solutions.

\subsection{Phishing Webpage Datasets}
\label{sec:Novelty_of_PhishOracle}
Numerous open-source phishing tools such as BlackEye~\citep{BlackEye}, ZPhisher~\citep{ZPhisher}, ShellPhish~\citep{ShellPhish}, etc. offer pre-designed phishing webpages tailored to a small number of specific brands. For example, Shellphish~\citep{ShellPhish} has a dataset containing pre-designed phishing webpages for $29$ distinct brands. Similarly, ZPhisher~\citep{ZPhisher} has pre-designed phishing webpages dataset for $33$ brands with each webpage incorporating external resources such as \texttt{CSS}, \texttt{PHP}, and \texttt{JS} files. Consequently, whenever there is a need for a phishing webpage related to any of these brands, the same webpage is consistently fetched from the dataset. These tools exhibit two limitations: a) they cannot \textit{generate} phishing webpages for brands not included in their predefined set, and b) because they already have pre-designed phishing webpages for each brand, they cannot provide a phishing webpage with diverse/new phishing features for the same brand. The tools are inherently \textit{static}, confined to a fixed set of brands.

\subsection{Phishing Webpage Detection Approaches}
\label{sec:ML_DL_and_LLM_Phishing_Detection_Approaches}
ML and DL techniques have significantly advanced phishing webpage detection by training models on large datasets containing phishing and legitimate samples. These datasets include URLs, webpage content, and screenshots. Phishing samples are obtained from repositories such as PhishTank\footnote{\label{PhishTank}PhishTank, \url{https://www.phishtank.com/index.php}} and OpenPhish\footnote{OpenPhish, \url{https://openphish.com/index.html}}, and legitimate samples are collected from Common Crawl\footnote{Common Crawl, \url{https://commoncrawl.org/}} and Stuff Gate\footnote{Stuff Gate, \url{https://stuffgate.com.websiteoutlook.com/}}. Alexa\footnote{Alexa Top Websites, \url{https://www.expireddomains.net/alexa-top-websites/}} and Tranco\citep{LePochat2019} provide a list of top-ranked legitimate domains, the content of which can be scraped. Datasets like UCI\footnote{UCI, \url{https://archive.ics.uci.edu/dataset/327/phishing+websites}}, provide feature-extracted files containing both legitimate and phishing samples, including URLs, webpage content, and third-party features. ML models use these features to identify patterns associated with phishing webpages, while DL techniques use neural networks to analyze more complex patterns, such as visual similarity in webpage screenshots and logos.

\subsubsection{ML and DL-based Phishing Webpage Detection Approaches}
\hfill \\
Li \textit{et al.}~\citep{li2019stacking} proposed a Stack model (GBDT, XGBoost and LightGBM) for phishing webpage detection using URL and HTML-based features. In this approach, Word2Vec model~\citep{Word2Vec} is used to learn HTML strings and represent them as vector encodings. The Stack model outperforms phishing detection solutions~\citep{zhang2007cantina, varshney2016phish, verma2015character} by achieving an accuracy of $96.45 \%$. The performance of the Stack model is better than individual ML classifiers for phishing webpage detection.

As DL techniques have evolved alongside the adoption of visual-similarity-based approaches in phishing webpage detection, Fu \textit{et al.}~\citep{fu2006detecting} proposed a visual similarity-based approach for phishing webpage detection. The algorithm consists of three stages: a) capturing a screenshot of the suspicious webpage, b) normalizing the screenshot to a fixed size, and c) describing the normalized image to a signature consisting of the pixel color and coordinate features. The Earth Movers' Distance (EMD) algorithm is used to compute the visual similarity between the obtained visual signature and the legitimate webpage signatures and classifies the suspicious webpage as phishing or legitimate. However, this approach performs the classification at the pixel level and does not consider the text. As a result, it cannot detect visually dissimilar webpages.

Afroz \textit{et al.}~\citep{afroz2011phishzoo} proposed PhishZoo, a visual-based phishing webpage detection approach based on the webpage profile. A profile of a webpage contains SSL, URL, HTML content and logo features. As a browser loads a webpage, its profile is generated and compared with the stored profiles of legitimate webpages. If the loaded webpage profile does not match with the legitimate SSL, URL, HTML content, and logos, then the user is warned of the webpage possibly being phishing. PhishZoo has advantages over URL-based approaches but also has limitations, such as the approach failing to classify a webpage containing previous logo versions, and if the logos are tilted over $30$ degrees.

Similarity-based phishing detection approaches, such as VisualPhishNet~\citep{abdelnabi2020visualphishnet}, identify phishing attempts by analyzing the visual similarity between webpage screenshots. VisualPhishNet utilizes a triplet convolutional neural network (CNN) trained to embed images into a feature space, where visually similar webpages are mapped closer together. The model compares a given webpage screenshot against a set of reference images from known legitimate brands. However, the approach has limitations: it may produce false positives (when different brands have visually similar webpage layouts) and false negatives (when a phishing webpage and its legitimate counterpart exhibit significant design variations despite targeting the same brand). Additionally, the reliance on visual similarity makes the system susceptible to adversarial attacks that modify key visual elements, while maintaining the phishing intent.

Lin \textit{et al.}~\citep{lin2021phishpedia} propose Phishpedia to detect phishing webpages and identify its target brand in a reference brand list. Phishpedia adopts a two-step approach: \textit{object detection} and \textit{brand identification}. The object in this context refers to the \textit{identity logo} present on a webpage screenshot for a given brand. In the first step, the region proposal network (RPN)~\citep{ren2015faster} analyzes the screenshot layout and predicts boxes around logos on the screenshot. The Faster RCNN model, based on the Detectron2 framework (from Facebook~\citep{wu2019detectron2}), selects the identity logo having the highest confidence score from the set of candidate logos. The next step is logo classification, wherein a ResNetV2 network~\citep{he2016identity} extracts diverse features from various logo variants belonging to the same brand. The classification task is trained on Logo2K+ dataset~\citep{wang2020logo}. This trained ResNetV2 model is linked with a global average pooling (GAP) layer to generate vectors that represent each logo. These vector representations are compared to logos in the reference brand list using cosine similarity~\citep{han2012getting}. The brand exhibiting the highest similarity score is identified as the target brand. Upon identifying the brand represented by a logo, the system verifies the legitimacy of the webpage by cross-referencing its domain with the domain associated with the identified brand. Phishpedia outperforms EMD~\citep{fu2006detecting}, PhishZoo~\citep{afroz2011phishzoo} and LogoSENSE~\citep{bozkir2020logosense} in terms of brand identification, phishing detection and runtime overhead.

Building on Phishpedia, Liu \textit{et al.}~\citep{liu2022inferring} propose PhishIntention, which enhances phishing detection by incorporating \textit{credential-taking} intention analysis alongside brand identification. Experiment results show that PhishIntention outperforms EMD~\citep{fu2006detecting}, VisualPhishNet~\citep{abdelnabi2020visualphishnet}, PhishZoo~\citep{afroz2011phishzoo}, and Phishpedia~\citep{lin2021phishpedia} in phishing detection, achieving higher precision while at similar recall. Its ability to extract and validate phishing intentions makes it a more effective solution against phishing attacks.

\subsubsection{LLM-based Approaches}
\hfill \\
With technological advancements, LLMs have gained attention in various sectors and their ability to extract contextual information from webpages has made them increasingly valuable in phishing detection. However, existing reference-based phishing detection solutions~\citep{lin2021phishpedia, liu2022inferring} are constrained by their reliance on reference brand lists, limiting their scalability to a large number of brands.
To address this, Li \textit{et al.}~\citep{li2024knowphish} propose an LLM-assisted approach that extracts brand information from HTML content and determines whether a webpage attempts to collect user credentials. By not relying solely on logos, this approach expands the phishing detection capability of existing reference-based solutions~\citep{lin2021phishpedia, liu2022inferring}. In a recent work, Lee \textit{et al.}~\citep{lee2024multimodal} introduce a two-stage system (requiring no training) that utilizes MLLMs (Gemini, Claude3, and GPT-4) for phishing webpage detection. In the first stage, the MLLM identifies the brand of the webpage by analyzing various aspects of the webpage (such as the HTML content and screenshot). In the second stage, an MLLM verifies whether the identified brand matches with the domain of the webpage. This approach outperforms the existing screenshot-based phishing webpage detection model VisualPhishNet~\citep{abdelnabi2020visualphishnet}.

MLLMs pretrained on massive datasets have the potential to assist or replace the existing models for phishing webpage detection, as they do not require labeled data for training and continuous maintenance (retraining), while still having the capability to understand webpage semantics. Therefore, we take a proactive approach and evaluate an LLM-based phishing detection approach~\citep{lee2024multimodal} against evasion attacks. 

\subsubsection{Robustness of Phishing Detection Models Against Adversarial Attacks}
\hfill \\
Recent works in phishing detection also concentrate on how the phishing detection models perform to classify adversarial phishing evasions. Sabir \textit{et al.}~\citep{sabir2022reliability} propose URLBUG, an adversarial URL generator designed to generate adversarial phishing URLs by targeting the domain, path and top-level domain parts of URLs. These generated adversarial URLs are used to evaluate the robustness of several ML-based phishing URL detectors. The experimental results reveal that these adversarial URLs significantly challenge the ML-based phishing URL detectors, showing vulnerabilities and limitations.

Object detection models are also prone to adversarial attacks, allowing attackers to create adversarial logos that closely resemble the legitimate ones and evade logo-based phishing webpage detectors~\citep{divakaran2022phishing}. Apruzzese \textit{et al.}~\citep{apruzzese2023real} examines the real-world impact of adversarial techniques on phishing webpage detection models. Their findings reveal that attackers use simple, low-complexity methods such as blurring, cropping, and masking logos to evade ML-based phishing webpage detectors. Another work by Lee \textit{et al.}~\citep{lee2023attacking} generates adversarial logos by adding noise to brand logos in a reference brand list, enabling these logos to evade logo-based phishing webpage detection solutions.

In another work, Apruzzese \textit{et al.}~\citep{apruzzese2022mitigating} introduce two sets of adversarial attacks: one where an attacker modifies only a few known features and another where the attacker has partial knowledge of the feature set. They develop a robust phishing webpage detection model that withstands these attacks, demonstrating significant performance improvements over existing phishing webpage detection models, which struggle against these adversarial attacks. Ji \textit{et al.}~\citep{ji2024evaluating} recently performed an extensive assessment of visual similarity-based phishing webpage detection models. The robustness of these phishing detection models is also evaluated against adversarial phishing webpages that utilize visual modifications like rotation, repositioning, resizing, and blurring, showing that even slight modifications can degrade the detection accuracy of the models. To improve the model's robustness, the authors argue to combine text recognition with visual analysis and use preprocessing methods like scaling and denoising to reduce the effects of adversarial alterations.

Inspired by recent efforts to evaluate the robustness of the phishing webpage detection models against adversarial phishing webpages~\citep{apruzzese2022mitigating, apruzzese2023real, lee2023attacking, ji2024evaluating}, we develop a tool that generates such adversarial phishing webpages. This tool aims to enhance phishing webpage detection models by identifying vulnerabilities through adversarial techniques, particularly targeting visual elements such as logos.

\section{Threat Model}
\label{sec:Threat_Model}
In this work, we consider a black-box adversary who aims to evade phishing webpage detection models by generating adversarial phishing webpages using widely observed visual transformation techniques, particularly logo manipulations.

The adversary generates adversarial phishing webpages by applying logo-based transformation techniques, which include adjusting opacity, applying Gaussian blur, and overlaying mesh/watermarks. These transformation techniques have been widely used in prior works~\citep{ji2024evaluating, apruzzese2023real, hao2024doesn} to evaluate the robustness of visual-based phishing webpage detection models.

The adversary’s goal is to generate adversarial phishing webpages targeting brands, such that the generated phishing webpages evade detection. We assume the adversary does not have access to deployed configurations or model weights. This setup aligns with a realistic black-box threat model. The adversary can use PhishOracle to modify the visual appearance of a legitimate webpage and generate adversarial variants embedded with different visual transformations.

Thus, the attack strategy is to download a legitimate webpage for a targeted brand, apply logo transformation techniques, generate adversarial phishing webpages, and evaluate the robustness of the phishing webpage detection models. A successful evasion occurs when the model predicts the brand on the webpage screenshot with low confidence, reducing the likelihood of detection.

Our target systems are phishing webpage detection models that rely on visual-based features to identify the brand in a webpage screenshot and compare it with the domain to detect phishing.

\textit{Out of Scope}: We assume attackers do not have access to deployed configurations or model parameters. As a result, advanced white-box attacks that require internal model access, such as gradient-based methods (e.g., FGSM~\citep{goodfellow2014explaining}, DeepFool~\citep{moosavi2016deepfool}), are not considered in our work. Similarly, logo perturbation attacks like Generative Adversarial Perturbations (GAP)~\citep{lee2023attacking}, although feasible in a black-box setting, are beyond the scope of our current work.

\section{Proposed Phishing Webpage Generation Tool: \texttt{PhishOracle}}
\label{sec:Proposed_Phishing_Webpage_Generation_Tool_PhishOracle}

Attackers create phishing webpages to collect users' sensitive information. Typically, they modify specific HTML tags like \texttt{<form>, <a>, <link>, <script>}, etc., redirecting the acquired sensitive information to databases or email addresses owned by the attackers. In addition to content-based webpage features, maintaining style consistency is another critical aspect that can deceive users. Typically, all webpages in the same organization adhere to a uniform style. The overall style similarity focuses on the visual aesthetics of a webpage, which can be characterized by various CSS design elements such as background colors, text alignments, font family, font size and subtle variations in favicons and logos. When integrated into a legitimate webpage, these visual-based features do not significantly alter its layout, but are sufficient to mislead users into believing that the phishing webpage originates from the legitimate domain owner. The visual resemblance between the generated phishing webpage and the legitimate styling is remarkably close, further deceiving users into perceiving the phishing webpage as a replica of the legitimate one. However, for these phishing webpages to reach users--the actual targets of the attackers--they must first evade phishing webpage detection models. To achieve this, adversarial phishing webpages are created with subtle modifications that allow them to evade phishing webpage detection models, increasing the likelihood that they will reach the users and deceive them.

\begin{figure}[t]
    \centering
    \begin{tikzpicture}[node distance=1.8cm]
        \node (start) [startstop] {Start};
        \node (in) [io, right of=start, xshift=3cm] {\shortstack{\textbf{Input:} \\ One Legitimate URL \\ Set of phishing features}};
        \node (pro1) [process, right of=in, xshift=5cm] {\shortstack{Request the URL, \\ Download the landing webpage}};
        \node (pro2) [process, below of=pro1] {\shortstack{Add randomly selected content \\ and visual-based phishing \\ features to the webpage}};
        \node (out) [io, below of=in] {\shortstack{\textbf{Output:} \\ Phishing webpage containing \\ set of embedded phishing \\ features}};
        \node (stop) [startstop, below of=start] {Stop};
    
        \draw [arrow] (start) -- (in);
        \draw [arrow] (in) -- (pro1);
        \draw [arrow] (pro1) -- (pro2);
        \draw [arrow] (pro2) -- (out);
        \draw [arrow] (out) -- (stop);
    \end{tikzpicture}
\caption{Phishing Webpage Generation Tool}
\label{fig:PhishOracle_Flowchart}
\Description{Phishing Webpage Generation Tool}
\end{figure}

\begin{table*}[th]
\caption{List of Phishing Features based on the Content and Visual Attributes in a Webpage}
\label{tab:Phishing_Features}
\begin{tabular}{ccll}
\toprule
\textbf{Type} & \textbf{Naming} & \textbf{Phishing Features} & \textbf{Explanation} \\ \midrule 
\multirow{27}{*}{\rotatebox[origin=c]{90}{\textbf{Content}}} & \textit{C1} & Hypertext reference & Updating \texttt{href} with: \texttt{href="\#"}, \texttt{"\#content"}, \texttt{"\#skip"} or \\ &&& \quad \texttt{"Javascript:void(0)"} \\ 
 
 & \textit{C2} & Disable key functions & Restrict users to view source code by disabling \textit{f11} and \\ &&& \quad \textit{Ctrl + U} \\ 
 
 & \textit{C3} & \texttt{href} lookalike characters & Alphabets in \texttt{href} value are replaced by lookalike characters \\ &&& \quad Ex: replacing \texttt{a} with a lookalike character {\texttt{ā, á, ä}}, etc. \\ 
 
 & \textit{C4} & Hide links appearing on & This feature does not allow users to view the \texttt{href} link \\ && \quad status bar & \quad even on hovering on the link \\
 
 & \textit{C5} & Disable anchor tags & Does not allow a user to navigate to other pages by \\ &&& \quad clicking on the anchor tags \\

 & \textit{C6} & Replace blank space & Blank spaces present in the container elements of HTML \\ && \quad  with a character & \quad like \texttt{h1, p, span} tags are replaced with characters \\ &&& \quad with \texttt{style=color: transparent;} \\

 & \textit{C7} & Save credentials & The credentials entered in the HTML form input tags are \\ &&& \quad stored in a local file \\

 & \textit{C8} & Disable other login & Webpages containing login buttons like Google, GitHub, \\ && \quad buttons & \quad  LinkedIn, etc are disabled and the credentials entered \\ &&& \quad in the visible login page are stored locally. \\
 
 & \textit{C9} & Pop-up Login & Clicking the login or sign-up button triggers a login page \\ &&& \quad  to appear. The action field in the \texttt{<form>} tag is then \\ &&& \quad altered, to store the credentials locally \\

 & \textit{C10} & Pop-up Login by & Clicking the anchor tags, opens up a login page \\ && \quad clicking on \texttt{a} tags & \quad and the credentials are stored locally \\

 & \textit{C11} & IFrame tag with & \texttt{<iframe>} tags are added with a login page, the credentials \\ && \quad login page & \quad entered are stored locally \\
 
 & \textit{C12} & Add dummy tags & Dummy \texttt{<img>}, \texttt{<link>}, \texttt{<script>}, \texttt{<a>} and \texttt{<div>} tags \\ &&& \quad are added to increase the DOM structure of webpage \\ \midrule

\multirow{11}{*}{\rotatebox[origin=c]{90}{\textbf{Visual}}} & \textit{V1} & Body Opacity & CSS \texttt{opacity} adjusts webpage transparency from $0$ to $1$. \\  

 & \textit{V2} & Text Styling & The \texttt{font-family} changes the style of webpage text \\
 
 & \textit{V3} & Opacity on Logo & CSS \texttt{opacity} property makes the logo image transparent. \\ &&& \texttt{<img>} tags containing images with \texttt{.png} and \texttt{.svg} \\ &&& \quad extensions are considered and a \texttt{opacity} of $0.1$ to $0.35$ is \\ &&& \quad added to make the logo transparent \\
 
 & \textit{V4} & Adding Watermark on logo & A watermark is added either on the bottom right corner or \\ &&& \quad diagonally on logo image \\

 & \textit{V5} & Image Transformations & Techniques like adding (1) clockwise or anti-clockwise \\ &&& \quad rotation, (2) Gaussian blur, (3) grey-mesh and (4) noise, \\ &&& \quad transforms the logo image
 \\ \bottomrule
\end{tabular}
\end{table*}

We propose \texttt{PhishOracle} to generate adversarial phishing webpages for any given legitimate webpage by randomly embedding content-based and visual-based features (refer Table~\ref{tab:Phishing_Features}). Figure~\ref{fig:PhishOracle_Flowchart} depicts the workflow of \texttt{PhishOracle}, with two inputs: a legitimate URL and a set/count of phishing features to add to the input legitimate webpage. Firstly, the legitimate URL is requested to fetch and save the webpage content and associated resources. Secondly, the algorithm parses the webpage content to deduce the list of applicable phishing features from Table~\ref{tab:Phishing_Features}. For example, if the webpage content contains anchor tags (\texttt{<a>}) then, relevant phishing features like \textit{C1}, \textit{C3}, \textit{C4}, \textit{C5} and \textit{C10} (refer Table~\ref{tab:Phishing_Features}) are embedded in the webpage to generate its corresponding phishing webpage.

\begin{figure*}[thbp]
    \centering
    \begin{subfigure}{0.49\textwidth}
        \centering
        \includegraphics[width=1.0\textwidth]{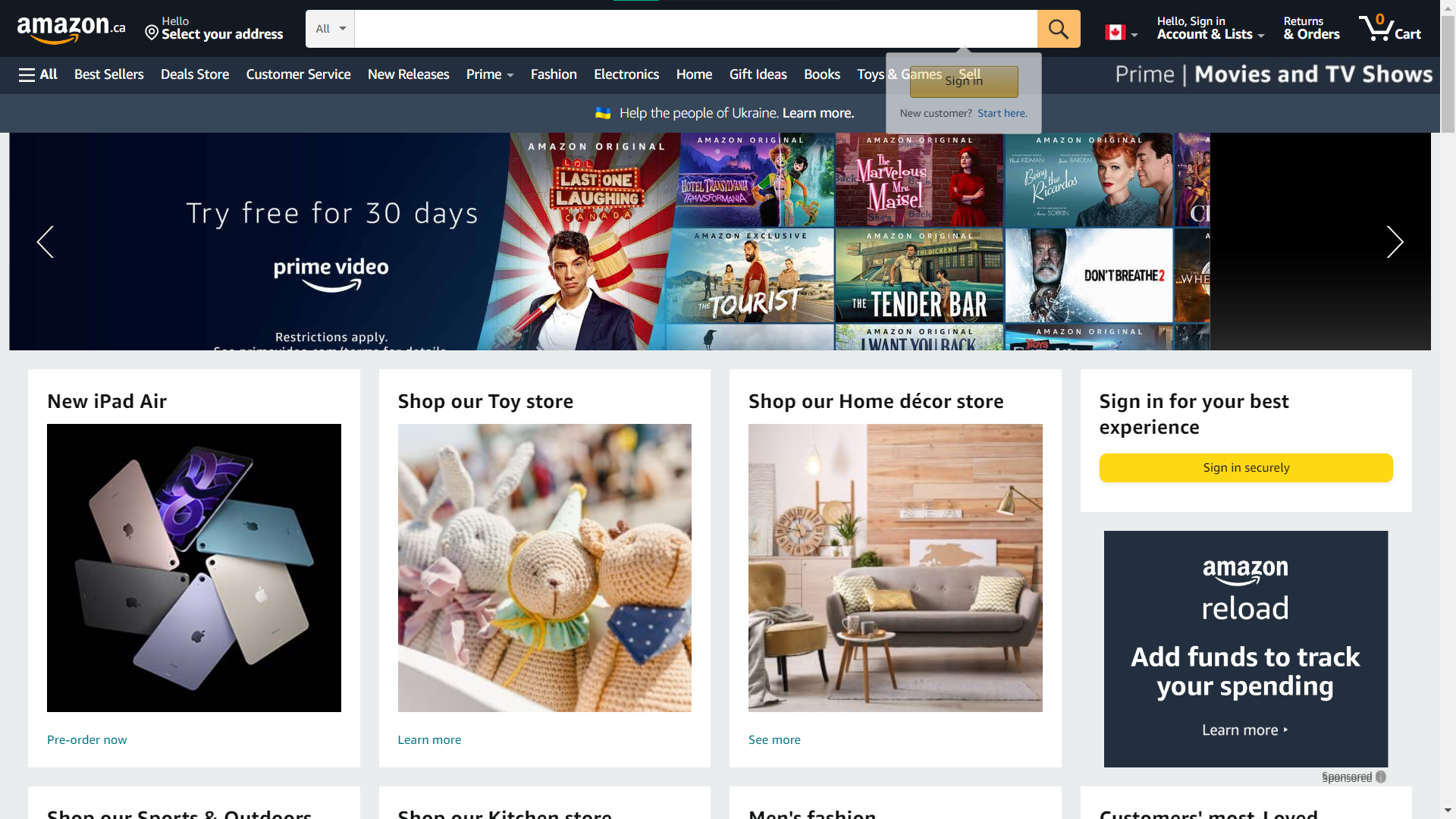}
        \caption{Legitimate webpage}
        \label{fig:considered_legitimate_webpage}
    \end{subfigure}
    \hfill
    \begin{subfigure}{0.49\textwidth}
        \centering
        \includegraphics[width=1.0\textwidth,]{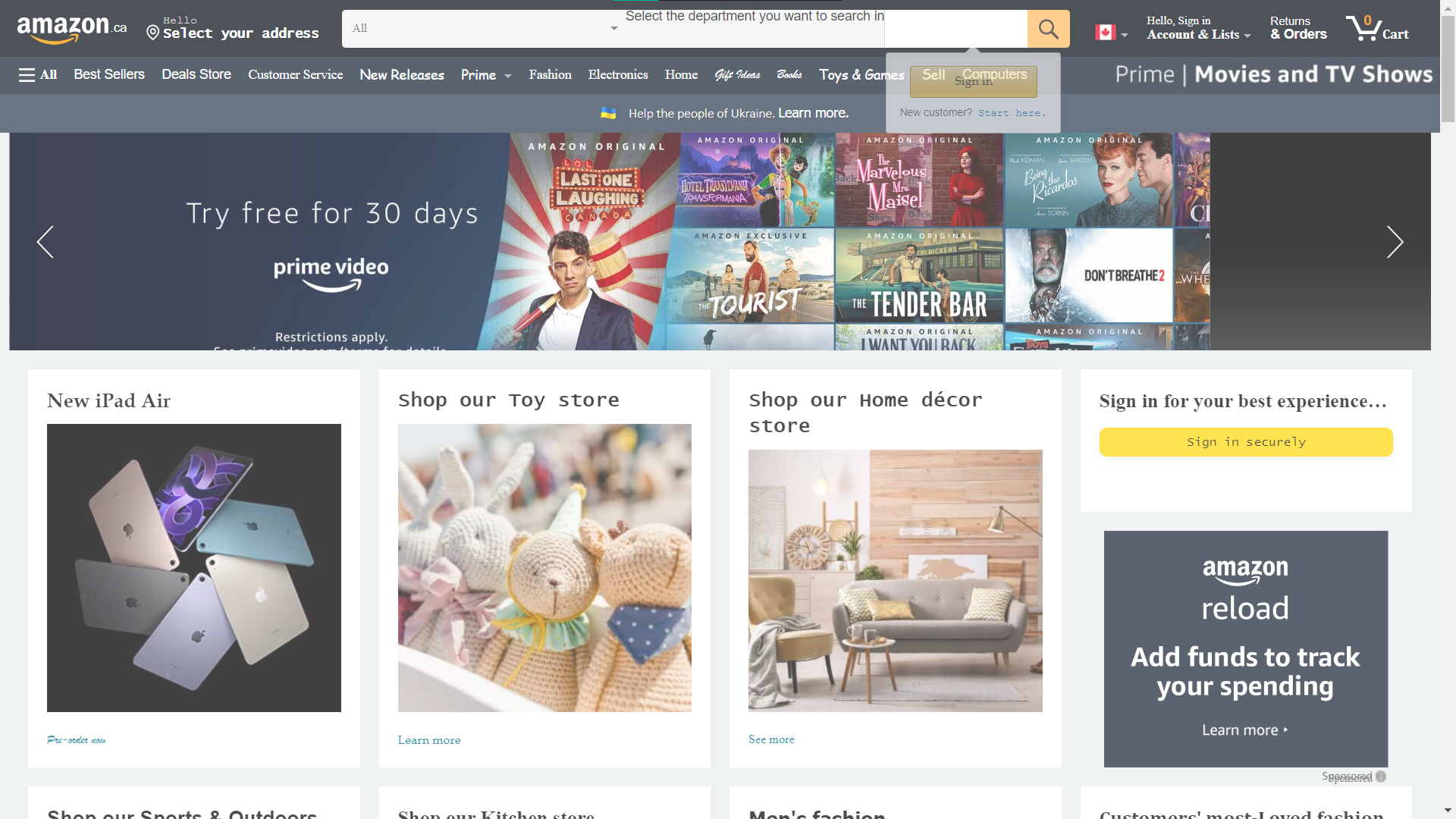}
        \caption{Generated Phishing webpage}
        \label{fig:generated_phishing_webpage}
    \end{subfigure}
    \caption{\texttt{PhishOracle}-generated webpage}
    \label{fig:PhishOracle_WebPage}
    \Description{Legitimate and \texttt{PhishOracle}-generated webpage}
\end{figure*}

Figure~\ref{fig:PhishOracle_WebPage} shows a legitimate webpage and its corresponding phishing webpage generated by \texttt{PhishOracle} by embedding content-based (\textit{C2}, \textit{C5}, \textit{C7}, \textit{C9}, and \textit{C12}) and visual-based (\textit{V1}, \textit{V2}) phishing features as listed in Table~\ref{tab:Phishing_Features}.

To produce a lookalike URL for the phishing webpage generated by \texttt{PhishOracle}, a combination of \textit{homoglyphs}, \textit{prefixes}, and \textit{suffixes} is utilized on the legitimate domain name. This combination results in the creation of domains that closely resemble legitimate domains, followed by corresponding URL parameters. For instance, if we consider the legitimate domain \texttt{facebook.com}, this process generates a lookalike domain such as \texttt{facebock-login.co}. A selection of homoglyphs examples includes substituting \texttt{d} with \texttt{cl}, \texttt{m} with \texttt{nn}, \texttt{w} with \texttt{vv}, \texttt{l} with \texttt{1}, \texttt{c} with \texttt{o}, \texttt{m} with \texttt{rn}, etc. Additionally, there are prefixes like \texttt{secure-}, \texttt{logon-}, and \texttt{login-}, as well as suffixes such as \texttt{-login}, \texttt{-logon}, and \texttt{-secure}, etc. to enhance the resemblance to legitimate domains.

\section{Performance Evaluation}
\label{sec:Experimental_Setup}
In this section, we outline the performance metrics used to evaluate phishing webpage detection models. We detail the clean dataset used for model training and the evasion dataset -- containing adversarial phishing webpages generated by \texttt{PhishOracle} -- to evaluate the robustness of these models. Additionally, we discuss the evaluation of security tools against these adversarial phishing webpages.

\subsection{Performance Metrics}
\label{sec:Performance_Metrics}
The fundamental metrics to evaluate classifiers are True Positive (\textit{TP}), False Positive (\textit{FP}), True Negative (\textit{TN}), and False Negative (\textit{FN}). In the context of the phishing detection models, these metrics are used for determining the count of actual to predicted values as shown in Table~\ref{tab:Confusion_Matrix}. The metrics \textit{TP, FP, TN} and \textit{FN} are elaborated as follows:
\begin{itemize}
    \item True Positive (\textit{TP}): The model correctly identifies a phishing website as phishing.
    \item False Positive (\textit{FP}): The model incorrectly identifies a legitimate website as phishing.
    \item True Negative (\textit{TN}): The model correctly identifies a legitimate website as legitimate.
    \item False Negative (\textit{FN}): The model incorrectly identifies a phishing website as legitimate.
\end{itemize}
\begin{table}[ht]
    \centering
    \caption{Confusion Matrix}
    \label{tab:Confusion_Matrix}
    \begin{tabular}{cccc}
        \toprule
        \multicolumn{2}{l}{\multirow{2}{*}{}} & \multicolumn{2}{c}{\textbf{Predicted}} \\ \cmidrule(l){3-4} 
        \multicolumn{2}{l}{} & \textit{Phishing} & \textit{Legitimate} \\ \cmidrule(r){1-2}
        \multirow{2}{*}{\makecell{\textbf{Actual}}} & \textit{Phishing} & True Positive & False Negative \\
         & \textit{Legitimate} & False Positive & True Negative \\ \bottomrule
    \end{tabular}
\end{table}
These metrics are used to compute the \textit{Accuracy ($\frac{TP+TN}{TP+FP+TN+FN}$), Precision ($\frac{TP}{TP+FP}$), Recall ($\frac{TP}{TP+FN}$}) and \textit{F1-Score ($\frac{2\times Precsion\times Recall}{Precision+Recall}$)} of classifiers.

\subsection{Dataset Description}
\label{sec:Dataset_Description}
With technological advancements, attackers find new features to evade existing phishing webpage detection solutions. Various solutions~\citep{zhang2007cantina, varshney2016phish, verma2015character, li2019stacking} consider URL and HTML features, and a few solutions~\citep{fu2006detecting, afroz2011phishzoo, abdelnabi2020visualphishnet, lin2021phishpedia} use visual similarity features to classify new phishing webpages. With the current research focus on evaluating the robustness of phishing webpage detection models against adversarial phishing webpages, we evaluate existing models for phishing webpage detection using adversarial phishing webpages generated by \texttt{PhishOracle}.

\begin{table}[ht]
    \centering
    \caption{Dataset Collection Summary}
    \label{tab:Collected_Dataset_Description}
    \begin{tabular}{cllrr}
        \toprule
        \multirow{2}{*}{\textbf{Notation}} & \multirow{2}{*}{\textbf{Description}} & \multirow{2}{*}{\textbf{Collection Period}} & \multicolumn{2}{c}{\textbf{Sample Size}} \\ \cmidrule{4-5}
        ~ & ~ & ~ & Phishing & Legitimate \\ \midrule
        \multirow{2}{*}{$\mathcal T^1$} & Contains URL, HTML content, and webpage & Dec'23 to Jan'24 & 9,105 & 8,727 \\ & screenshots of phishing and legitimate samples \\ \midrule
        \multirow{4}{*}{$\mathcal T^2$} & Contains URL, HTML content, and webpage & May'24 & $0$ & 9,067 \\ & screenshot of legitimate samples, with manually \\ & annotated logos required to train the Faster \\ & RCNN model in the Phishpedia experiment \\ \bottomrule
    \end{tabular}
\end{table}

\begin{table}[th]
    \centering
    \caption{\textit{CleanSet}, and \textit{EvasionSet} Description for Training and Evaluating the Phishing Webpage Detection Models}
    \label{tab:CleanSet_and_EvasionSet_Description}
    \begin{tabular}{llrrr}
    \toprule
        \multirow{2.5}{*}{\textbf{Notation}} & \multirow{2.5}{*}{\textbf{Description}} & \multicolumn{3}{c}{\textbf{Sample Size}} \\ \cmidrule{3-5}
        ~ & ~ & Phishing & Legitimate & \texttt{PhishOracle} \\ &&&& (generated phishing) \\ \midrule
        \multirow{3}{*}{\textit{CleanSet\textsuperscript{1}}} & $\mathcal T^1$ dataset used to train the Stack model~\citep{li2019stacking} & 9,105 & 8,727 & $0$ \\ & and to evaluate the LLM-based phishing \\ &  detector~\citep{lee2024multimodal} \\ \midrule
        \multirow{8}{*}{\textit{CleanSet\textsuperscript{2}}} & $\mathcal T^2$ dataset used to train the Faster RCNN & $909$ & 9,067 & 0 \\ & model, and $\sim$$900$ phishing ($\subset \mathcal T^1$), and $82$ \\ & legitimate samples ($\subset \mathcal T^2$), are used to \\ & evaluate the performance of brand \\ & identification and phishing detection model \\ & of the entire Phishpedia system~\citep{lin2021phishpedia}, and \\ & the LLM-based phishing detector~\citep{lee2024multimodal} \\ 
        \midrule
        \multirow{6}{*}{\textit{CleanSet\textsuperscript{3}}} & $\sim$$900$ phishing samples ($\subset\mathcal T^1$), and $\sim$$600$ & $896$ & $911$ & 0 \\ & legitimate samples covering the $41$ target \\ & brands + $\sim$$300$ legitimate webpage \\ & screenshots ($\subset \mathcal T^2$), from non-target brands, \\ & used to train the triplet CNN model in the \\ & VisualPhishNet model~\citep{abdelnabi2020visualphishnet} \\ \midrule
        \multirow{6}{*}{\textit{EvasionSet\textsuperscript{1}}} & Contains adversarial phishing webpages & $0$ & 1,000 & 1,000 \\ & generated by \texttt{PhishOracle} for legitimate \\ & webpages selected from $\mathcal T^1$ dataset to \\ & evaluate the Stack model~\citep{li2019stacking}, and \\ & LLM-based phishing detector~\citep{lee2024multimodal} \\ 
        \midrule
        \multirow{7}{*}{\textit{EvasionSet\textsuperscript{2}}} & Contains $170$ adversarial phishing webpages & 0 & $82$ & $170$ \\ & generated by \texttt{PhishOracle} for $82$ \\ & legitimate webpages ($\subset T^2$ dataset) in \\ & the reference brand list of Phishpedia~\citep{lin2021phishpedia} \\ & and to evaluate the performance of the \\ & Phishpedia~\citep{lin2021phishpedia} and the LLM-based \\ & phishing detector~\citep{lee2024multimodal} \\ \midrule
        \multirow{4}{*}{\textit{EvasionSet\textsuperscript{3}}} & Contains $41$ legitimate and $107$ corresponding & $0$ & $41$ & $107$ \\ & adversarial phishing webpages ($\subset $ \text{} \textit{EvasionSet\textsuperscript{2}}), \\ & and used to evaluate the performance of \\ & VisualPhishNet~\citep{abdelnabi2020visualphishnet} \\ \bottomrule
    \end{tabular}
\end{table}

In our study, we select the Stack model~\citep{li2019stacking} due to its effectiveness in ML-based phishing webpage detection, to assess its capability to classify \texttt{PhishOracle}-generated adversarial phishing webpages that include a variety of phishing features. Next, we select VisualPhishNet~\citep{abdelnabi2020visualphishnet}, a screenshot-based phishing webpage detection model that analyzes the visual similarity between webpage screenshots to distinguish phishing webpages and legitimate ones by learning discriminative features based on brand identity and layout structure. Moreover, we select Phishpedia~\citep{lin2021phishpedia}, known for its robustness in phishing webpage detection and identifying target brands in webpage screenshots. This makes it ideal for identifying brands in adversarial phishing webpage screenshots generated by \texttt{PhishOracle}, incorporating different logo transformation techniques. Finally, we select an LLM-based phishing detector~\citep{lee2024multimodal}, selecting Gemini~1.5 Flash~\citep{Gemini_Flash} as the LLM, to evaluate the solution's capability in identifying the target brand in HTML content and webpage screenshots of \texttt{PhishOracle}-generated adversarial phishing webpages, and verify whether the identified brand matches with the domain of the webpage.

As the datasets to train these selected models~\citep{li2019stacking, abdelnabi2020visualphishnet, lin2021phishpedia} are outdated, we retrain them using our latest datasets (see Table~\ref{tab:Collected_Dataset_Description}) and evaluate their robustness in detecting the adversarial phishing webpages generated by \texttt{PhishOracle}. We collect these dataset samples by using a Python script that includes various libraries, such as \texttt{Selenium}, \texttt{requests}, and \texttt{urllib.parse}. This script scraped 9,105 phishing samples from PhishTank\footref{PhishTank}, and 8,727 legitimate samples from Tranco~\citep{LePochat2019}, collected between December $2023$ and January $2024$. This collected dataset is referred to as $\mathcal T^1$ in Table~\ref{tab:Collected_Dataset_Description}. However, some brands in the reference brand list of the Phishpedia experiment prohibited downloading the webpage content but allowed capturing screenshots, while others prohibited both. To ensure their inclusion in the dataset, we manually captured screenshots of these websites. As a result, we collected a total of 9,067 legitimate webpage screenshots, referred to as $\mathcal T^2$ in Table~\ref{tab:Collected_Dataset_Description}. Furthermore, we manually annotated the brand logos in each of the 9,067 webpage screenshots to train the Faster RCNN model for brand identification in the Phishpedia experiment.

The collected samples in Table~\ref{tab:Collected_Dataset_Description} are categorized into \textit{CleanSet}, and \textit{EvasionSet}, each serving a distinct purpose for training and evaluating the robustness of the selected phishing webpage detection models. Table~\ref{tab:CleanSet_and_EvasionSet_Description} describes the categories of these \textit{CleanSet}, and \textit{EvasionSet} datasets. \textit{CleanSet\textsuperscript{1}} corresponds to the $\mathcal T^1$ dataset (refer Table~\ref{tab:Collected_Dataset_Description}), which is used to train the Stack model~\citep{li2019stacking} and to evaluate LLM-based phishing detector~\citep{lee2024multimodal}. It includes 9,105 phishing samples and 8,727 legitimate samples, both containing URL, HTML content, and webpage screenshots. \textit{CleanSet\textsuperscript{2}} corresponds to the $\mathcal T^2$ dataset (refer Table~\ref{tab:Collected_Dataset_Description}), primarily used to train the Faster RCNN model for brand identification in the Phishpedia experiment~\citep{lin2021phishpedia}. It consists of 9,067 legitimate samples. Moreover, $909$ phishing samples ($\subset \mathcal T^1$ dataset) and $82$ legitimate samples ($\subset \mathcal T^2$ dataset), are specifically used to evaluate the performance of the brand identification and phishing detection of the entire Phishpedia system~\citep{lin2021phishpedia}, as well as the LLM-based phishing detector~\citep{lee2024multimodal}. \textit{CleanSet\textsuperscript{3}} is used to train the VisualPhishNet model~\citep{abdelnabi2020visualphishnet}. It consists of $896$ phishing samples ($\subset \mathcal T^1$ dataset) covering $41$ target brands, $\sim$$600$ legitimate webpage screenshots, and $\sim$$300$ additional legitimate webpage screenshots (from non-target brands $\subset \mathcal T^2$ dataset). This dataset is used to train the triplet CNN model for visual similarity-based phishing detection.

The \textit{EvasionSet} datasets in Table~\ref{tab:CleanSet_and_EvasionSet_Description}---containing adversarial phishing webpages generated by \texttt{PhishOracle}---evaluate the trained models. \textit{EvasionSet\textsuperscript{1}} contains 1,000 legitimate samples ($\subset \mathcal T^1$ dataset) with corresponding 1,000 adversarial phishing webpages generated using \texttt{PhishOracle}, and is used to evaluate the Stack model~\citep{li2019stacking}, and the LLM-based phishing detector~\citep{lee2024multimodal} (referred as \textit{LLM-PD\textsuperscript{H}} in Table~\ref{tab:Performance_Comparison_Stack_Model_Phishpedia_Gemini_Pro_Vision}). \textit{EvasionSet\textsuperscript{2}} consists of $82$ legitimate samples ($\subset \mathcal T^2$ dataset), selected from Phishpedia’s reference brand list, along with $170$ adversarial phishing webpages generated using \texttt{PhishOracle}. This dataset is used to evaluate the Phishpedia system~\citep{lin2021phishpedia}, as well as the LLM-based phishing detector~\citep{lee2024multimodal} (referred as \textit{LLM-PD\textsuperscript{S}} in Table~\ref{tab:Performance_Comparison_Stack_Model_Phishpedia_Gemini_Pro_Vision}). \textit{EvasionSet\textsuperscript{3}} contains $107$ phishing and $41$ legitimate samples, which are a subset of \textit{EvasionSet\textsuperscript{2}} and focus on the $41$ target brands in VisualPhishNet~\citep{abdelnabi2020visualphishnet}. This dataset is used to assess the robustness of the VisualPhishNet model against adversarial phishing webpages generated by \texttt{PhishOracle}.


\subsection{Generating \texttt{PhishOracle} Validation Dataset}
\label{sec:Generating_PhishOracle_Validation_Dataset}
Developed using Python version $3.10.2$, our \texttt{PhishOracle} tool uses various libraries, including \texttt{BeautifulSoup}, \texttt{requests}, \texttt{webbrowser}, \texttt{PIL}, and \texttt{urllib.parse}. For the Stack model experiment, we select 1,000 legitimate webpages ($\subset \mathcal T^1$ dataset) and embed relevant content-based and visual-based phishing features (refer Table~\ref{tab:Phishing_Features}) to generate corresponding 1,000 phishing webpages, forming the \textit{EvasionSet\textsuperscript{1}} dataset. The algorithm includes a mechanism for detecting relevant tags within the webpages. The detection of such tags triggers the embedding of corresponding phishing features into the webpage content. For instance, if the algorithm identifies \texttt{<a>} tag within the legitimate webpage, it embeds features such as \textit{C1}, \textit{C3}, \textit{C5}, \textit{C10} and \textit{C12} (refer Table~\ref{tab:Phishing_Features}), resulting in the generation of the corresponding phishing webpage. This entire process ensures the creation of balanced sets comprising legitimate and corresponding phishing webpages.

For VisualPhishNet, Phishpedia, and LLM-based phishing detection experiments, we select $82$ legitimate webpages ($\subset \mathcal T^2$ dataset), all belonging to Phishpedia's reference brand list. We then generate $170$ corresponding adversarial phishing webpages by incorporating visual-based features on logos (refer Table~\ref{tab:Phishing_Features}). Note that a single webpage can have multiple \texttt{PhishOracle}-generated phishing webpages, each generated by embedding different visual-based features, specifically by incorporating logo transformation techniques (refer to visual-based features in Table~\ref{tab:Phishing_Features}). The \textit{EvasionSet\textsuperscript{2}} dataset comprises screenshots of these $82$ legitimate and $170$ generated phishing webpages, which are used to evaluate Phishpedia and LLM-based phishing detector. A subset of \textit{EvasionSet\textsuperscript{2}}, referred to as \textit{EvasionSet\textsuperscript{3}}, is constructed by selecting phishing and legitimate samples associated with the $41$ target brands of VisualPhishNet. This dataset is specifically designed to assess the robustness of the VisualPhishNet model against phishing attempts targeting protected brands. Furthermore, we generate lookalike URLs for each of these generated phishing webpages (in \textit{EvasionSet\textsuperscript{1}}, \textit{EvasionSet\textsuperscript{2}}, and \textit{EvasionSet\textsuperscript{3}}) using homoglyphs, suffixes, and prefixes. The source code and the datasets used in this experiment are available on our GitHub repository~\citep{PHISHORACLE_GITHUB}.

\begin{figure}[th]
    \centering
    \includegraphics[width=0.55\linewidth]{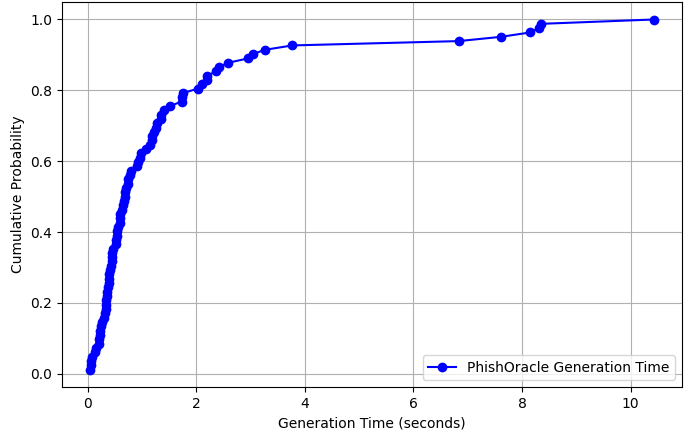}
    \caption{CDF of \texttt{PhishOracle} processing time for generating adversarial phishing webpages in the \textit{EvasionSet\textsuperscript{2}} dataset}
    \label{fig:PhishOracle_CDF}
    \Description{CDF of \texttt{PhishOracle} processing time for generating adversarial phishing webpages in the \textit{EvasionSet\textsuperscript{2}}.}
\end{figure}

Figure~\ref{fig:PhishOracle_CDF} shows the performance evaluation of \texttt{PhishOracle} on \textit{EvasionSet\textsuperscript{2}} dataset. The average processing time taken to embed a diverse set of content and visual-based phishing features into legitimate webpages is $\sim$$1.48$ seconds. The majority of the webpages are processed under $3$ seconds, with only a few outliers extending up to $10$ seconds due to potentially complex webpage structures. The CDF plot indicates that around $80 \%$ of webpages were processed in under $3$ seconds, demonstrating the efficiency of \texttt{PhishOracle} to generate adversarial phishing webpages.

\subsection{Validating Phishing Webpage Detection Models}
\label{sec:Validating_Phishing_Detection_Approaches}
In this section, we discuss the experiment setup for Stack model~\citep{li2019stacking}, VisualPhishNet~\citep{abdelnabi2020visualphishnet}, Phishpedia~\citep{lin2021phishpedia} and LLM-based phishing detector~\citep{lee2024multimodal}, and illustrate their performance on detecting adversarial phishing webpages generated by \texttt{PhishOracle}.

\subsubsection{Stack Model} \hfill \\
Li \textit{et al.}~\citep{li2019stacking} propose a Stack model for phishing webpage detection using URL and HTML features, without considering third-party-based features (like DNS records, web traffic, etc.).

\textbf{Setup:} To carry out our experiments, we use the feature extraction code of Stack Model made available by Phishpedia's authors on their GitHub repository~\citep{Phishpedia_GitHub} and we develop the Stack model using GBDT, XGBoost and LightGBM classifiers.

\textbf{Training \& Testing the model:} We use \textit{CleanSet\textsuperscript{1}} containing URLs and HTML contents of 9,105 phishing and 8,727 legitimate webpages; and the train:test split is set to 80:20. In our experiments, Stack model achieves a high detection rate (recall) of $\sim$$98.32 \%$ at a high precision of $\sim$$98.67 \%$ on the\textit{CleanSet\textsuperscript{1}} dataset.

\textbf{Validation on} \textit{EvasionSet\textsuperscript{1}}: For evaluating the robustness of the Stack model, we now use the \textit{EvasionSet\textsuperscript{1}} that contains URLs and HTML content of legitimate and \texttt{PhishOracle}-generated phishing webpages. Stack model's detection rate falls to $\sim$$70.86 \%$ for a comparable precision of $\sim$$98.61 \%$. Observe that, this is a significant reduction in comparison to the performance achieved on the test set \textit{CleanSet\textsuperscript{1}} (as mentioned above)---$\sim$$28 \%$ in recall for about the same precision. Therefore, the Stack model is not robust to simple feature modifications (refer Section~\ref{sec:Proposed_Phishing_Webpage_Generation_Tool_PhishOracle}), and confirms the observations made in a recent study~\citep{lee2020building}.

\begin{figure*}[ht]
    \centering
    \begin{subfigure}{0.49\textwidth}
        \centering
        \includegraphics[width=1.0\textwidth]{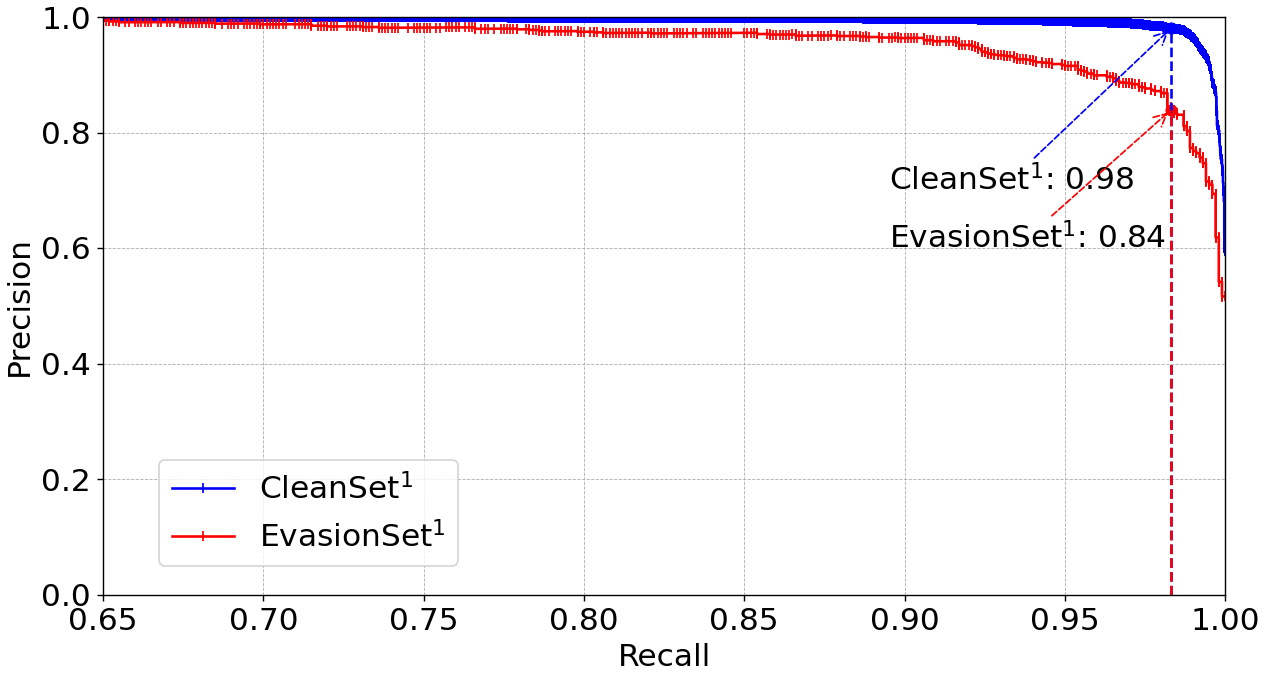}
        \caption{Precision vs Recall}
        \label{fig:Precision_vs_Recall_Stack_Model}
    \end{subfigure}
    \hfill
    \begin{subfigure}{0.49\textwidth}
        \centering
        \includegraphics[width=1.0\textwidth,]{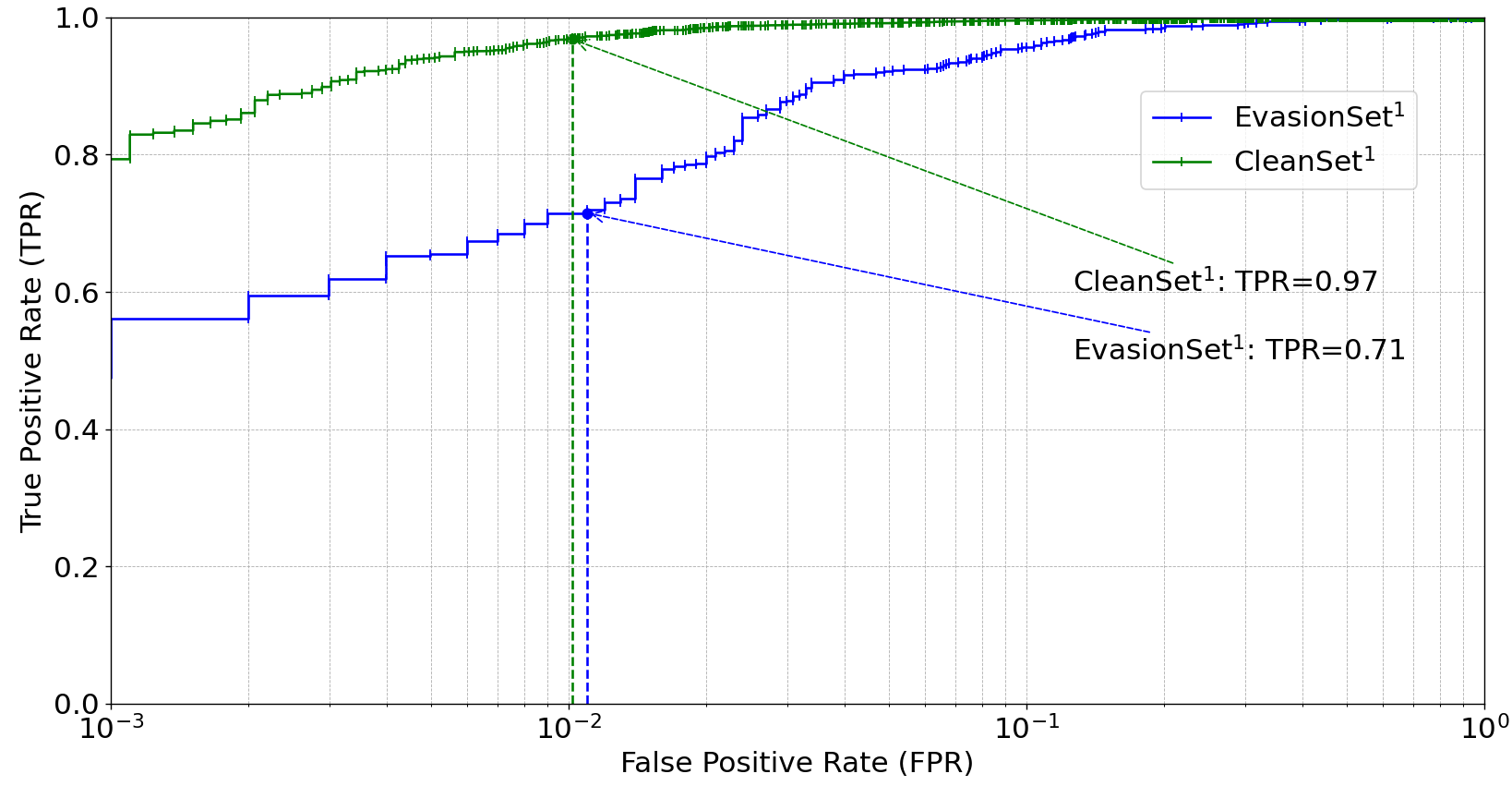}
        \caption{TPR vs FPR}
        \label{fig:TPR_vs_FPR_Stack_Model}
    \end{subfigure}
    \caption{Precision vs Recall (Figure~\ref{fig:Precision_vs_Recall_Stack_Model}) and TPR vs FPR (Figure~\ref{fig:TPR_vs_FPR_Stack_Model}) for the Stack Model}
    \label{fig:Stack_Model_Graphs}
    \Description{Precision vs. Recall and TPR vs FPR for the Stack Model}
\end{figure*}

Figure~\ref{fig:Precision_vs_Recall_Stack_Model} displays the precision-recall curve, illustrating the detection efficacy of the Stack model on the \textit{CleanSet\textsuperscript{1}} and \textit{EvasionSet\textsuperscript{1}} datasets. For a recall of $\sim$$98 \%$, the Stack model's precision drops significantly on the \textit{EvasionSet\textsuperscript{1}} to  $\sim$$84\%$ in contrast to $\sim$$98 \%$ on the \textit{CleanSet\textsuperscript{1}}. This performance reduction indicates the model's reduced effectiveness in detecting adversarial phishing webpages generated by \texttt{PhishOracle}. In addition, Figure~\ref{fig:TPR_vs_FPR_Stack_Model} displays the ROC curve for both datasets, drawn on a logarithmic scale to show the model's performance at low false positive rates (FPR). For a fixed FPR of $\sim$$10^{-2}$, the model achieves a true positive rate (TPR) of $\sim$$97\%$ on the \textit{CleanSet\textsuperscript{1}}, which drops to $\sim$$71\%$ on the \textit{EvasionSet\textsuperscript{1}}, indicating reductions in both precision and TPR for adversarial phishing webpages.

\subsubsection{VisualPhishNet} \hfill \\
Abdelnabi \textit{et al.}~\citep{abdelnabi2020visualphishnet} propose VisualPhishNet to detect phishing webpages by analyzing the visual similarity between webpage screenshots using a triplet CNN model.

\textbf{Setup}: We initiate the experiment by cloning the baseline model from the Phishpedia GitHub repository~\citep{Phishpedia_GitHub}.

\textbf{Training and Testing Model}: In this experiment, we use phishing and legitimate webpage screenshot datasets. The \textit{CleanSet\textsuperscript{3}} dataset consists of $\sim$$900$ phishing samples ($\subset \mathcal T^1$ dataset) targeting $41$ brands, along with $\sim$$600$ legitimate webpage screenshots for these $41$ target brands, collected by visiting their official website hyperlinks. Additionally, to ensure an unbiased training process, we include $\sim$$300$ legitimate webpage screenshots from non-target brands ($\subset \mathcal T^2$ dataset). This dataset is used to train the triplet CNN model in VisualPhishNet for visual similarity-based phishing detection. We follow the 60:40 train-test split, as suggested in \citep{abdelnabi2020visualphishnet}, to evaluate the model’s performance effectively. The retrained model achieves a precision of $\sim$$98.42\%$ at a recall of $\sim$$88.35\%$ on the \textit{CleanSet\textsuperscript{3}}.

\begin{figure*}[ht]
    \centering
    \begin{subfigure}{0.49\textwidth}
        \centering
        \includegraphics[width=1.0\textwidth]{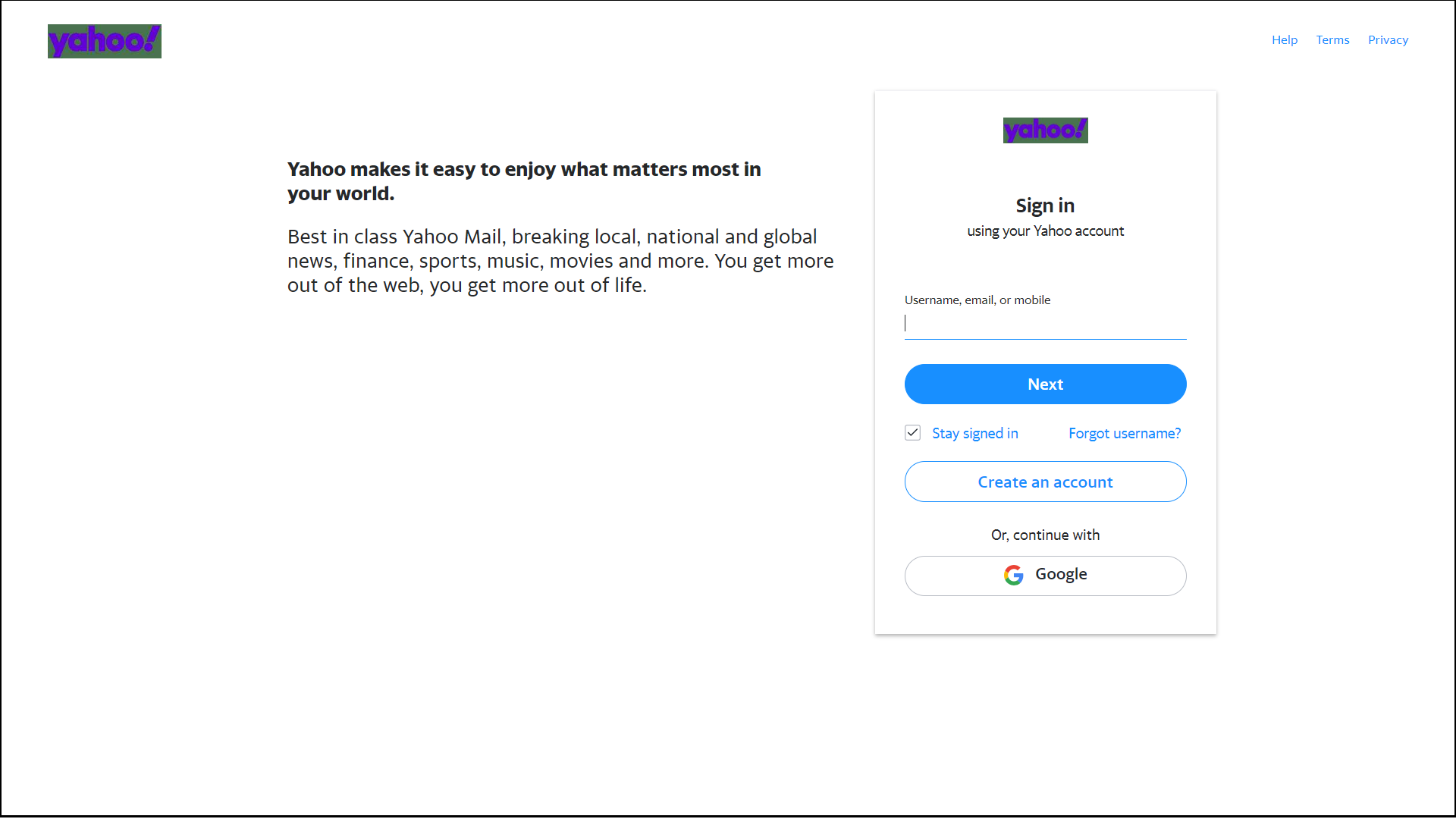}
        \caption{Adversarial Phishing Webpage targeting Yahoo}
        \label{fig:Adversarial_Phishing_Webpage_Targeting_Yahoo}
    \end{subfigure}
    \hfill
    \begin{subfigure}{0.49\textwidth}
        \centering
        \includegraphics[width=1.0\textwidth,]{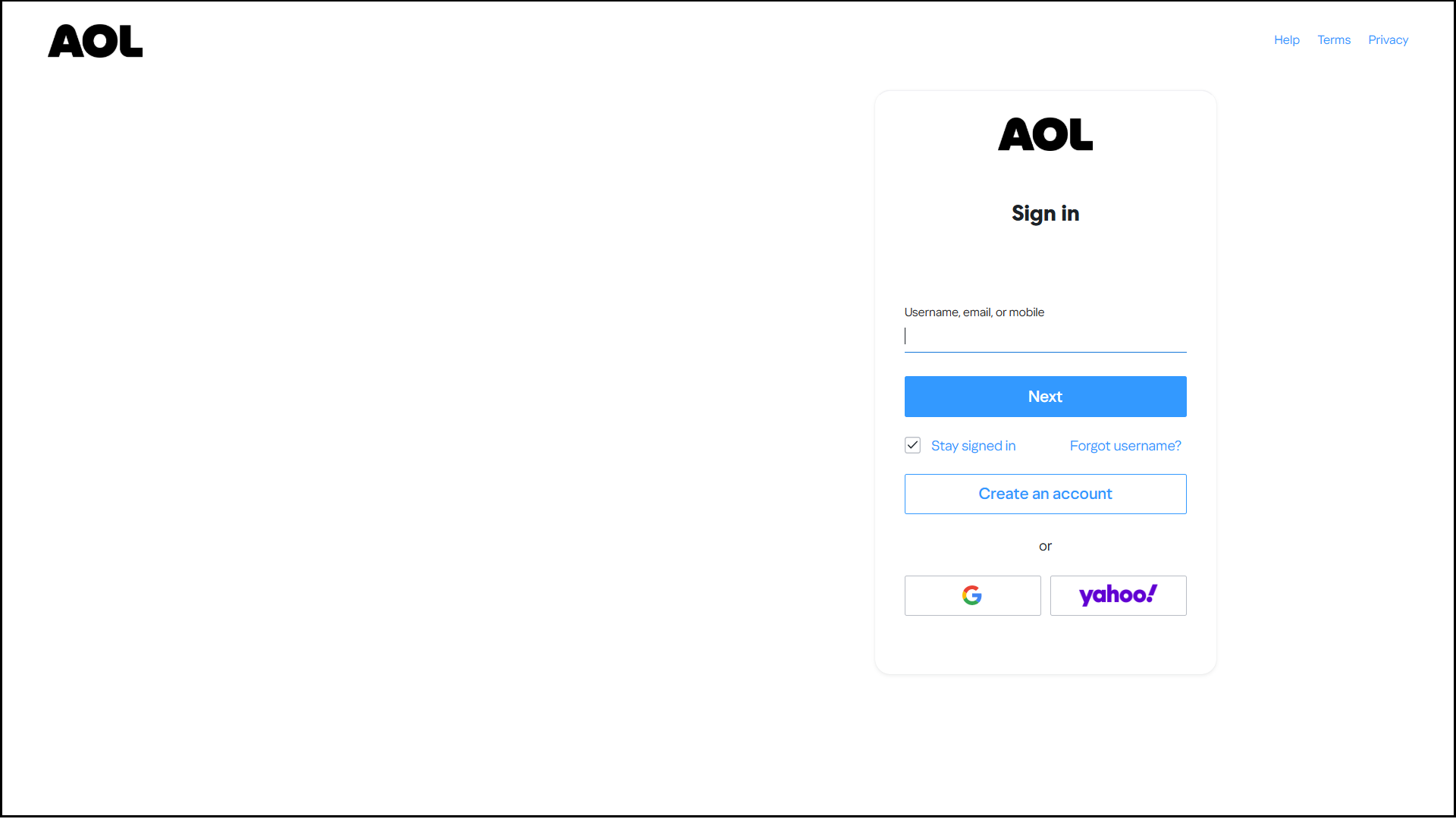}
        \caption{Legitimate Webpage of AOL}
        \label{fig:AOL_Legitimate_Webpage}
    \end{subfigure}
    \caption{\centering VisualPhishNet misclassify adversarial phishing webpage targeting Yahoo (Figure~\ref{fig:Adversarial_Phishing_Webpage_Targeting_Yahoo}) as AOL (Figure~\ref{fig:AOL_Legitimate_Webpage}) due to high similarity in their visual appearance}
    \label{fig:VisualPhishNet_Misclassification}
    \Description{VisualPhishNet misclassify an adversarial phishing webpage to another brand due to high similarity in their visual appearance}
\end{figure*}

\textbf{Validation on} \textit{EvasionSet\textsuperscript{3}}: Finally, we evaluate the trained VisualPhishNet model on the \textit{EvasionSet\textsuperscript{3}}, containing $41$ legitimate and $107$ adversarial phishing webpages generated by \texttt{PhishOracle}. This results in a considerably lower phishing detection rate 
(recall) of $\sim$$72.89\%$ at a 
precision of $\sim$$98.73 \%$. This demonstrates that the adversarial phishing webpages generated by \texttt{PhishOracle}, which incorporate logo transformation techniques (such as watermark, rotation, opacity, etc.), evade the VisualPhishNet model. For example, as shown in Figure~\ref{fig:VisualPhishNet_Misclassification}, the model incorrectly identifies an adversarial phishing webpage targeting Yahoo (Figure~\ref{fig:Adversarial_Phishing_Webpage_Targeting_Yahoo}) as AOL (Figure~\ref{fig:AOL_Legitimate_Webpage}), due to similarities in the visual appearance of the webpage layouts, including logo placement and form fields.

\textbf{Observations}: To evaluate the robustness of VisualPhishNet, we use \textit{EvasionSet\textsuperscript{3}}, containing adversarial phishing webpages generated by \texttt{PhishOracle}. The detection rate of VisualPhishNet drops to $\sim$$72.89 \%$ for a comparable high precision of $\sim$$98.73 \%$ (refer Table~\ref{tab:Performance_Comparison_Stack_Model_Phishpedia_Gemini_Pro_Vision}). Observe that this is a significant reduction compared to its performance on the test set \textit{CleanSet\textsuperscript{3}} (as mentioned above)--$\sim$$16 \%$ decrease in recall for about the same precision. This indicates that VisualPhishNet is affected by the logo-based adversarial manipulations. These findings align with recent evaluations~\citep{ji2024evaluating}, showing that the screenshot-based models like VisualPhishNet struggle with accurate brand identification when logos are altered, removed or replaced.

\subsubsection{Phishpedia} \hfill \\
Lin \textit{et al.}~\citep{lin2021phishpedia} propose Phishpedia to detect phishing webpages targeting a specific brand in a reference brand list.

\textbf{Setup:} We initiate the experiment by cloning the Phishpedia GitHub repository~\citep{Phishpedia_GitHub}. To carry out this experiment on Phishpedia, we manually labelled identity logos on 9,067 legitimate webpage screenshots in the $\mathcal T^2$ dataset.

\textbf{Training \& Testing the model:} In this experiment, we consider two datasets. For evaluating the object detection task, we use $\mathcal T^2$ dataset (described in Table~\ref{tab:Collected_Dataset_Description}), containing 9,067 legitimate URLs and webpage screenshots, and split it into 7,539 samples for training and 1,528 for testing. This results in a train:test split of 83:17. For the brand identification task, which uses the Siamese model pretrained on Logo2K+ dataset~\citep{wang2020logo}, and the phishing detection task, we use \textit{CleanSet\textsuperscript{2}}, containing $909$ phishing webpage screenshots with manually labelled identity logos and $82$ legitimate webpage screenshots from the reference brand list, with a train:test split of 80:20.

\textbf{Results:} The logo detection task achieves the mean Average Precision (mAP)~\citep{henderson2017end} of $\sim$$43.42 \%$, which is considered good in object detection. The brand identification model accurately detects the brand on $\sim$$97.77 \%$ of phishing webpages. In phishing detection, the entire Phishpedia system achieves a precision of $\sim$$97.25 \%$ at a recall of $\sim$$81.63 \%$. The values are slightly lower than the precision of $\sim$$98.2 \%$ at a recall of $\sim$$87.1 \%$ reported in the paper~\citep{lin2021phishpedia}; the reasons could be that, we have a smaller dataset for training in our work, or there is a difference in the quality of data. However, close to $\sim$$81.63 \%$ recall at a high precision of $\sim$$98 \%$ is still a good performance.

\textbf{Validation on} \textit{EvasionSet\textsuperscript{2}}: Finally, we evaluate the newly trained Phishpedia system on the \textit{EvasionSet\textsuperscript{2}}, containing $82$ legitimate webpage screenshots and $170$ \texttt{PhishOracle}-generated adversarial phishing webpage screenshots. This results in a precision of $\sim$$76.40 \%$ at a recall of $\sim$$40 \%$ for phishing detection (refer Table~\ref{tab:Performance_Comparison_Stack_Model_Phishpedia_Gemini_Pro_Vision}). Observe that the \texttt{PhishOracle}-generated adversarial phishing webpage screenshots containing logo transformation techniques (such as adding watermark, noise, etc.), can evade brand identification in the Phishpedia model.

\subsubsection{Multimodal LLMs} \hfill \\
With the emergence of LLMs, we today have several MLLMs that can process, analyze and generate texts, images and other multimedia contents. The capabilities of the MLLMs to analyze texts and images make them suitable for our task of brand identification in HTML content and webpage screenshots. In a recent work, Lee \textit{et al.}~\citep{lee2024multimodal} propose a two-stage system of MLLMs for phishing webpage detection. In the first stage, an MLLM analyzes various aspects of webpage, including its HTML content and screenshots, to identify the brand. In the second stage, an MLLM verifies whether the identified brand matches the domain of the webpage (URL), flagging mismatches as phishing. The approach employs separate prompts for analyzing HTML content and webpage screenshots to identify the brand, along with an additional prompt to verify whether the identified brand matches the domain of the webpage. In our study, we adopt these prompts to evaluate the LLM-based phishing detection pipeline~\citep{lee2024multimodal} using Gemini~1.5 Flash~\citep{Gemini_Flash} on both clean and adversarial phishing webpages, using \textit{CleanSet\textsuperscript{1}} and \textit{CleanSet\textsuperscript{2}}; and \textit{EvasionSet\textsuperscript{1}} and \textit{EvasionSet\textsuperscript{2}}, respectively.

To ensure a fair comparison with existing phishing detection models, we utilize two Gemini-based variants: \textit{LLM-PD\textsuperscript{H}} and \textit{LLM-PD\textsuperscript{S}} from~\citep{lee2024multimodal}. The Stack model, which relies on URL and HTML-based features, is evaluated on \textit{EvasionSet\textsuperscript{1}}. Accordingly, we use the HTML prompt from~\citep{lee2024multimodal} and provide the corresponding HTML contents as input to the LLM-based pipeline, referring to this variant as \textit{LLM-PD\textsuperscript{H}}. Similarly, Phishpedia, which identifies brands in webpage screenshots is evaluated on \textit{EvasionSet\textsuperscript{3}} -- so we use the screenshot prompt from~\citep{lee2024multimodal} and input the corresponding 
 webpage screenshots into the LLM-based pipleine, referring to this as \textit{LLM-PD\textsuperscript{S}}. This approach ensures that the performance of LLM-based approach is directly compared to state-of-the-art phishing webpage detection models on the same evasion datasets.
\begin{enumerate}
    \item Brand identification from \textbf{HTML content}

    \hspace{0.5cm} \textbf{Setup}: This experiment evaluates the ability of the LLM-based pipeline from~\citep{lee2024multimodal} using  Gemini~\citep{Gemini_Flash} to identify brands from the HTML content of webpages. We employ the HTML-based prompt from \citep{lee2024multimodal} and refer to this variant as \textit{LLM-PD\textsuperscript{H}} (as shown in Table~\ref{tab:Performance_Comparison_Stack_Model_Phishpedia_Gemini_Pro_Vision}). The model is tested on the \textit{CleanSet\textsuperscript{1}} dataset, which comprises phishing and legitimate samples. We evaluate the performance of \textit{LLM-PD\textsuperscript{H}} in brand identification and verifying its consistency with the domain of the webpage. The model achieves a precision of $\sim$$76.41 \%$ at a recall of $\sim$$73.83 \%$, and an F1-Score of $\sim$$75.09 \%$.

    \hspace{0.5cm} \textbf{Robustness Evaluation on} \textit{EvasionSet\textsuperscript{1}}: To assess the robustness of \textit{LLM-PD\textsuperscript{H}} against adversarial phishing webpages, we evaluate its performance on \textit{EvasionSet\textsuperscript{1}}, which contains phishing samples generated by \texttt{PhishOracle}.

    \hspace{0.5cm} \textbf{Results}: On \textit{EvasionSet\textsuperscript{1}} dataset, \textit{LLM-PD\textsuperscript{H}} achieves a precision of $\sim$$85.16 \%$ at a recall of $\sim$$67.55 \%$, and an F1-Score of $\sim$$75.34 \%$.

    \hspace{0.5cm} \textbf{Observations}: \textit{LLM-PD\textsuperscript{H}} achieves a precision of $\sim$$85.16 \%$ at a recall of $\sim$$67.55 \%$, and an F1-Score of $\sim$$75.34 \%$ on \textit{EvasionSet\textsuperscript{1}}, indicating a decline in recall compared to the \textit{CleanSet\textsuperscript{1}}. This indicates an increase in false negatives, where more phishing webpages are misclassified as legitimate. However, this corresponding increase in precision offsets this drop, resulting in a comparable F1-Score. These results suggest that the LLM-based phishing detection pipeline remains relatively robust against adversarial phishing attacks, though further improvements may be necessary to enhance recall.

    \item Brand identification from \textbf{webpage screenshots}

    \hspace{0.5cm} \textbf{Setup}: In this experiment, we evaluate the ability the LLM-based phishing pipeline~\citep{lee2024multimodal} to identify brands from webpage screenshots. We use the screenshot-based prompt from \citep{lee2024multimodal} and refer to this variant of Gemini as \textit{LLM-PD\textsuperscript{S}} (as shown in Table~\ref{tab:Performance_Comparison_Stack_Model_Phishpedia_Gemini_Pro_Vision}). The model is tested on the \textit{CleanSet\textsuperscript{2}}, a dataset consisting of both legitimate and phishing webpage screenshots. We analyze how effectively \textit{LLM-PD\textsuperscript{S}} identifies brand names and verifies them against the domain of the webpage. The model achieves an impressive precision of $\sim$$98.83 \%$ at a recall of $\sim$$97.25 \%$, and an F1-Score of $\sim$$98.03 \%$.

    \hspace{0.5cm} \textbf{Robustness Evaluation on} \textit{EvasionSet\textsuperscript{2}}: To assess the robustness of \textit{LLM-PD\textsuperscript{S}}, we evaluate its performance on \textit{EvasionSet\textsuperscript{2}}, which contains screenshots of legitimate webpages along with their corresponding adversarial phishing webpages variants generated by \texttt{PhishOracle}.

    \hspace{0.5cm} \textbf{Results}: Evaluation of \textit{LLM-PD\textsuperscript{S}} on \textit{EvasionSet\textsuperscript{2}} results in a precision of $100 \%$ at a recall of $\sim$$79.51 \%$, with an F1-Score of $\sim$$88.59 \%$.

    \begin{figure}[th]
        \centering
        \includegraphics[width=0.95\textwidth]{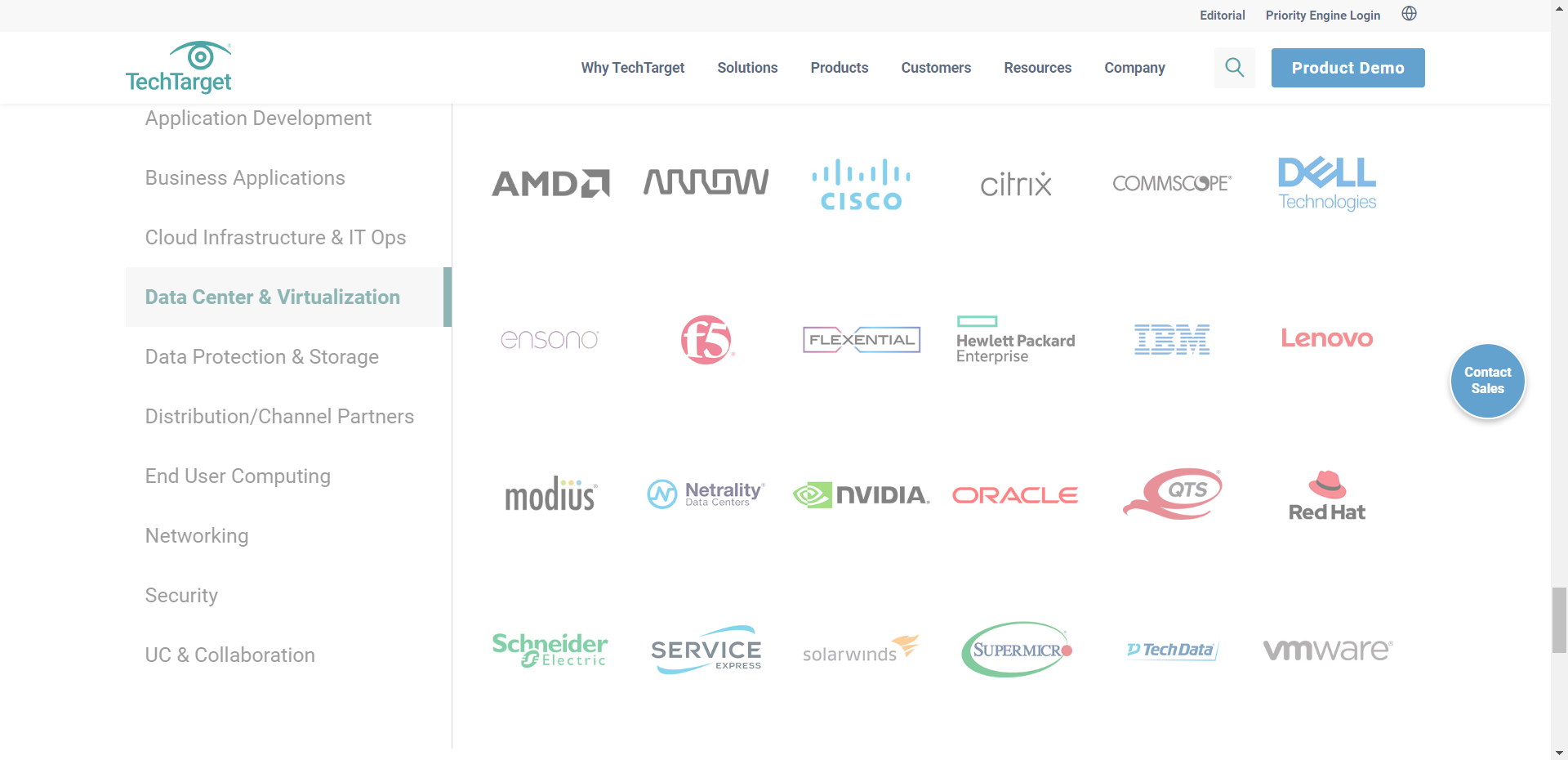}
        \caption{\centering Webpage Screenshot containing multiple logos of different brands}
        \label{fig:Multiple_Logos_Gemini_Pro_Vision}
        \Description{Webpage Screenshot containing multiple logos of different brands}
    \end{figure}

    \hspace{0.5cm} \textbf{Observations}: \textit{LLM-PD\textsuperscript{S}} is robust on the adversarial phishing webpages generated by \texttt{PhishOracle} but it experiences a drop of $\sim$$10 \%$ F1-Score on \textit{EvasionSet\textsuperscript{2}} as compared to \textit{CleanSet\textsuperscript{2}}. For instance, in an adversarial phishing webpage targeting Yahoo (see Figure~\ref{fig:Adversarial_Phishing_Webpage_Targeting_Yahoo}), the first-stage MLLM correctly identifies the brand and provides supporting evidence. The second-stage MLLM while comparing the identified brand with the domain name / URL (\texttt{http://signup-yahoo.com})  incorrectly associates it with the identified brand as belonging to the same company, leading to evasion. Nevertheless, \textit{LLM-PD\textsuperscript{S}} outperforms VisualPhishNet~\citep{abdelnabi2020visualphishnet} and Phishpedia~\citep{lin2021phishpedia} in effectively identifying brands on adversarial phishing webpages generated by \texttt{PhishOracle}. Furthermore, \textit{LLM-PD\textsuperscript{S}} efficiently identify the actual brand name in webpage screenshots containing logos of multiple brands. As an illustration, it correctly identifies the brand name as `TechTarget' in the webpage screenshot containing logos of multiple brands, as shown in Figure~\ref{fig:Multiple_Logos_Gemini_Pro_Vision}. Once the brand is correctly identified, comparing the URL's domain name with the identified brand's domain name gives the detection result. 
    
\end{enumerate}

\begin{table}[ht]
    \centering
    \caption{Performance of Stack Model, VisualPhishNet, Phishpedia, and \textit{LLM-PD\textsuperscript{H}}, and \textit{LLM-PD\textsuperscript{S}}}
    \label{tab:Performance_Comparison_Stack_Model_Phishpedia_Gemini_Pro_Vision}
    \begin{tabular}{lcrrrcrrr}
        \toprule
        \multirow{2}{*}{\textbf{Model}} & \multicolumn{4}{c}{\textit{CleanSet}} & \multicolumn{4}{c}{\textit{EvasionSet}} \\ \cmidrule(r){2-5} \cmidrule(l){6-9}
        & \textbf{Dataset} & \textit{Precision} & \textit{Recall} & \textit{F1-Score} & \textbf{Dataset} & \textit{Precision} & \textit{Recall} & \textit{F1-Score} \\ \midrule
        Stack Model~\citep{li2019stacking} & \textit{CleanSet\textsuperscript{1}} & $98.67 \%$ & $98.32 \%$ & $98.49 \%$ & \textit{EvasionSet\textsuperscript{1}} & $98.61 \%$ & $70.86 \%$ & $82.46 \%$ \\
        VisualPhishNet~\citep{abdelnabi2020visualphishnet} & \textit{CleanSet\textsuperscript{3}} & $98.42 \%$ & $88.35 \%$ & $93.11 \%$ & \textit{EvasionSet\textsuperscript{3}} & $98.73 \%$ & $72.89 \%$ & $83.87 \%$ \\
        Phishpedia~\citep{lin2021phishpedia} & \textit{CleanSet\textsuperscript{2}} & $97.25 \%$ & $81.63 \%$ & $88.76 \%$ & \textit{EvasionSet\textsuperscript{2}} & $76.40 \%$ & $40 \%$ & $52.51 \%$ \\
        \textit{LLM-PD\textsuperscript{H}} & \textit{CleanSet\textsuperscript{1}} & $76.41 \%$ & $73.83 \%$ & $75.09 \%$ & \textit{EvasionSet\textsuperscript{1}} & $85.16 \%$ & $67.55 \%$ & $75.34 \%$ \\
        \textit{LLM-PD\textsuperscript{S}} & \textit{CleanSet\textsuperscript{2}} & $98.83 \%$ & $97.25 \%$ & $98.03 \%$ & \textit{EvasionSet\textsuperscript{2}} & $100 \%$ & $79.51 \%$ & $88.59 \%$ \\ \bottomrule
    \end{tabular}
\end{table}

Table~\ref{tab:Performance_Comparison_Stack_Model_Phishpedia_Gemini_Pro_Vision} provides a comprehensive comparison of the Stack model, VisualPhishNet, Phishpedia, and \textit{LLM-PD\textsuperscript{H}}, and \textit{LLM-PD\textsuperscript{S}} in phishing webpage detection.

\takeaways{
\\ \indent i) The detection rate of the Stack model~\citep{li2019stacking} drops by around $28 \%$ on \texttt{PhishOracle}-generated \textit{EvasionSet\textsuperscript{1}} compared to its performance on \textit{CleanSet\textsuperscript{1}}, when evaluated at a comparable precision. 
\\ \indent ii) VisualPhishNet~\citep{abdelnabi2020visualphishnet} experiences a drop of around $16 \%$ in detection rate when evaluated on \textit{EvasionSet\textsuperscript{3}} relative to its performance on \textit{CleanSet\textsuperscript{3}}, when tested on a comparable precision. 
\\ \indent iii) Phishpedia~\citep{lin2021phishpedia} suffers a drastic drop of approximately $40 \%$ in detection rate on \textit{EvasionSet\textsuperscript{2}} due to adversarial logo transformations (refer to Table~\ref{tab:Phishing_Features}), at a $20 \%$ lower  precision. This indicates that transformed logos can evade Phishpedia. Another contributing factor could be the smaller training dataset used in our work. 
\\ \indent iv) Compared to other phishing detectors, LLM-based phishing detector (\textit{LLM-PD\textsuperscript{S}}) (with no task-specific training) is more robust against adversarial phishing webpages generated by \texttt{PhishOracle}. Nonetheless, it also experiences a decline of performance---around $10 \%$ in F1-Score  on \textit{EvasionSet\textsuperscript{2}} compared to \textit{CleanSet\textsuperscript{2}}.
}

\subsection{Evaluation of Security Tools against Adversarial Phishing Attacks}
\label{sec:Evaluation_of_Adversarial_Phishing_Webpages_Detection_and_Blacklisting_by_Security_Tools}
In this section, we present a $9$-day evaluation of adversarial phishing webpages generated by \texttt{PhishOracle}, focusing on their detection by security tools.

\begin{figure}[ht]
    \centering
    \includegraphics[width=0.95\linewidth]{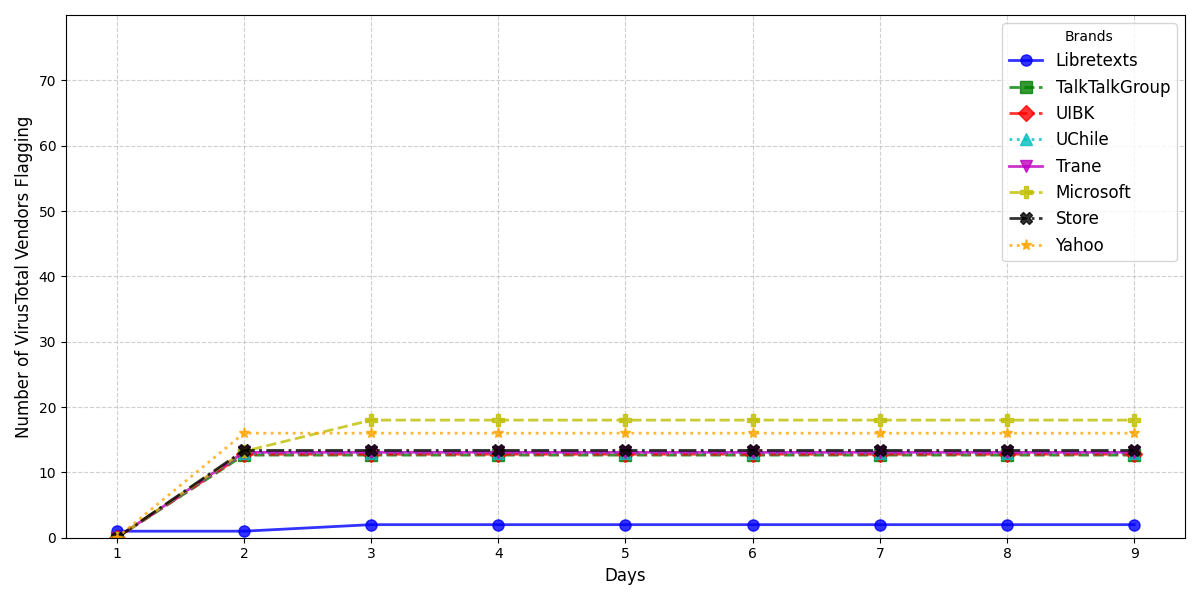}
    \caption{Evaluation of Security Tools Over $9$ Days Against Adversarial Phishing Attacks}
    \label{fig:9_Days_Analysis}
    \Description{Evaluation of Security Tools Over $9$ Days Against Adversarial Phishing Attacks}
\end{figure}

\textbf{Setup}: We conducted a 9-day study to evaluate effectiveness of the security tools in detecting the adversarial phishing webpages generated by \texttt{PhishOracle}. The evaluation focused on VirusTotal~\citep{Virustotal}, with more than $90$ security vendors, along with GSB and Microsoft Defender SmartScreen. We initiate the experiment by generating one adversarial phishing webpage for eight brands, including Microsoft, Yahoo, Store Steam, TalkTalk Group, UIBK, UChile, Trane, and LibreTexts. These webpages are designed by embedding diverse content-based features -- such as altering form action fields, embedding pop-up login forms, using Javascript-based navigation prevention to restrict user movement -- and visual-based features, including opacity, and logo transformation techniques (refer Table~\ref{tab:Phishing_Features}). Once generated, these adversarial phishing webpages are hosted on GitHub Pages, allowing them to be publicly accessible for evaluation.

\textbf{Evaluation Methodology}: The adversarial phishing webpages generated by \texttt{PhishOracle} were examined over a $9$-day span at $24$-hour intervals. The total number of security vendors in VirusTotal flagging these webpages as phishing or suspicious were documented. Additionally, the identification by GSB and Microsoft Defender SmartScreen were also tracked to observe how promptly these popular security solutions recognize these adversarial phishing webpages.

\textbf{Results}: With a total of $96$ security vendors on VirusTotal, the detection rates varied significantly among different brands. On Day $1$, only one phishing webpage targeting LibreTexts, was flagged by only one security vendor (Trustwave). The other phishing websites were still not marked by any of the security vendors. On Day $2$, the Microsoft phishing webpage was flagged by $6$ security vendors on VirusTotal, and the Microsoft Defender SmartScreen marked it as ``Dangerous Site''. Several other phishing webpages targeting brands such as TalkTalk Group, UIBK, UChile, Trane, Steam, and Yahoo were flagged by $13$ vendors, such as GSB, Trustwave, Webroot, and CyRadar. The phishing webpage from LibreTexts was marked as ``Dangerous'' by GSB. On Day $3$, the number of detections rose for phishing webpages targeting popular brands such as Microsoft and Yahoo, which were flagged by $18$ and $16$ security vendors respectively. Moreover, the LibreTexts phishing webpage was marked as ``Dangerous'' by Microsoft Defender SmartScreen, however, it remained unflagged by most of the security vendors on VirusTotal. From Days $4$ to $9$, the count of vendors marking these hosted adversarial phishing webpages remained stable. Figure~\ref{fig:9_Days_Analysis} displays the detection rates of security vendors against adversarial phishing webpages generated by \texttt{PhishOracle}.

\textbf{Observations}: The 9-day evaluation study showed that phishing webpages targeting well-known brands (e.g., Microsoft, Yahoo) were flagged by more security vendors than those targeting lesser-known brands (e.g., LibreTexts). However, detections were alarmingly slow, with the majority of security vendors failing to identify phishing webpages within the first $24$ hours. For instance, the Microsoft phishing webpage was flagged by only six vendors by Day $2$, meaning over $90 \%$ of the $96$ VirusTotal security vendors failed to detect it within the first day. Similarly, the LibreTexts phishing webpage remained largely undetected, with just one vendor identifying it by Day $1$.

These results align with prior research~\citep{lee20257}, which reported an average phishing site lifespan of $2.25$ days, showing that phishing webpages remain active long enough to deceive victims. Detection rates in our study followed a similar trend, where most phishing webpages were flagged only after two or more days, and detections stagnated after Day $3$. This delay is particularly concerning because even a few hours of undetected activity provides ample time for attackers to successfully execute phishing campaigns.

Our findings further support previous work~\citep{choo2023large}, which highlights inconsistencies in how VirusTotal vendors classify phishing URLs. While some vendors flagged phishing webpages within two days, most security solutions failed to detect them within the first $24$ hours, exposing critical detection gaps. Even for high-profile brands like Microsoft and Yahoo, only $16$ to $18$ out of $96$ security vendors flagged them by Day $3$, meaning $\sim$$80 \%$ of security vendors still failed to detect them by this time. This detection gap left phishing webpages accessible to potential victims for an extended period.

\section{User study and \texttt{PhishOracle} Web app}
\label{sec:User_Study_and_Contributions}
In this section, we discuss a user study to verify whether the phishing webpages generated by \texttt{PhishOracle} truly deceive the actual users. Moreover, we describe our \texttt{PhishOracle} web app, which is designed for both user education and phishing research purposes.

\subsection{User Study}
\label{sec:User_Study}
In this section, we assess the effectiveness of adversarial phishing webpages generated by \texttt{PhishOracle} in deceiving users. We conduct the study to evaluate the ability of users to determine the legitimacy of a webpage based solely on its visual appearance, without relying on the URL.

\textbf{Participants}: The user study aims to determine whether \texttt{PhishOracle}-generated adversarial phishing webpages can deceive users based on their visual appearance alone, even if detection models can be evaded. To achieve this, we selected $52$ participants within the age range of $19$ and $26$, consisting of academic students and IT professionals. A recent study~\citep{draganovic2023users} examined whether phishing webpages that evade the phishing webpage detection models can still deceive users based on visual appearance alone, which closely aligns with our research objective. Furthermore, the study suggests conducting a brief user study by presenting a limited number of phishing webpage samples to the participants for evaluation. This selection allowed us to assess whether IT professionals -- who are generally more knowledgeable about phishing threats and frequently encounter suspicious emails -- can analyze a webpage's visual appearance to determine its legitimacy. For academic students, the study also served as a form of user education, emphasizing the importance of inspecting webpage visuals for phishing indicators. We adhered to the ethical guidelines~\citep{finn2007designing} to safeguard user information confidentiality.

\begin{figure*}[th]
    \centering
    \begin{subfigure}{0.49\textwidth}
        \centering
        \includegraphics[width=1.0\textwidth]{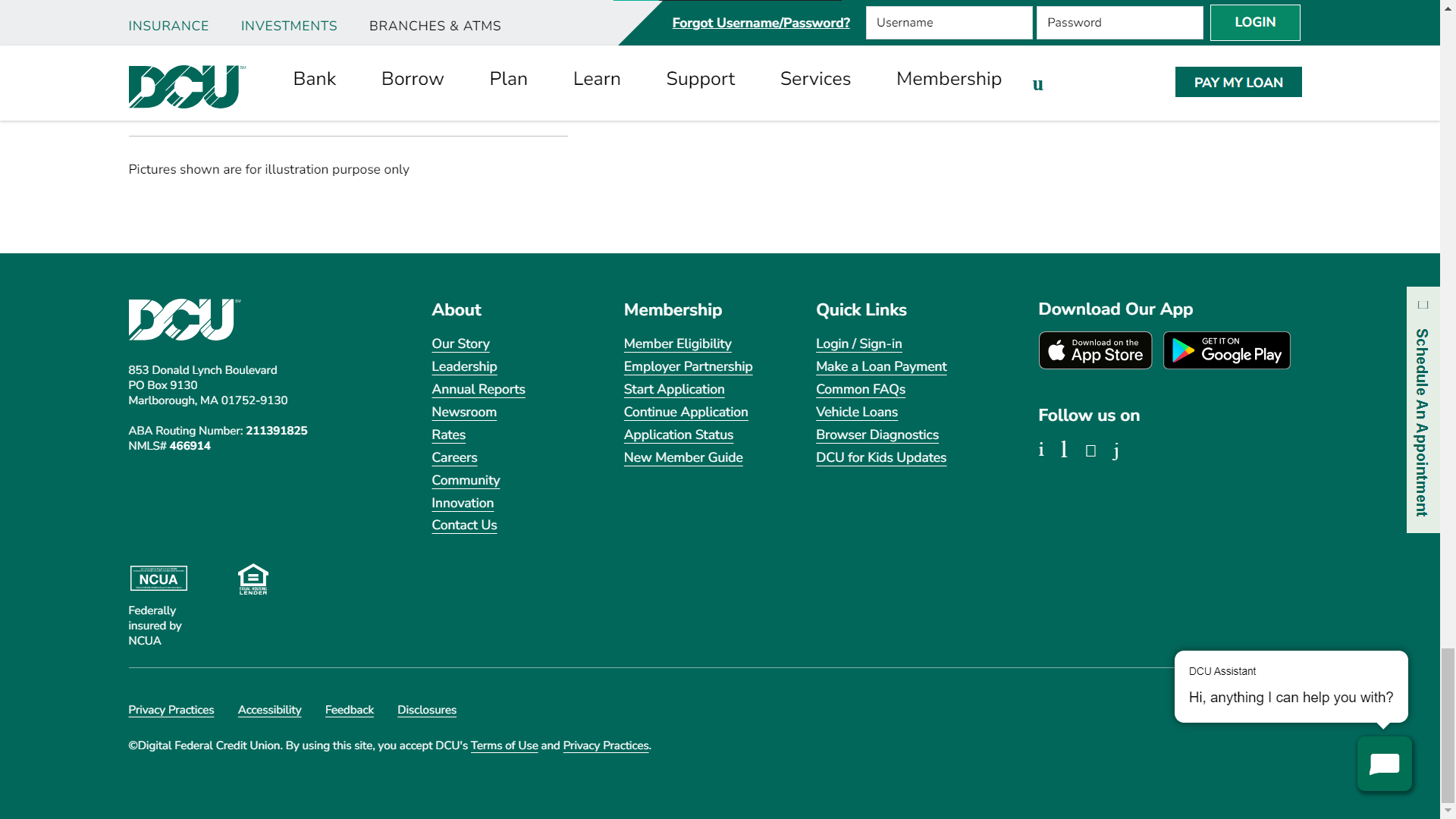}
        \caption{Legitimate webpage}
        \label{fig:considered_legitimate_webpage_user_study}
    \end{subfigure}
    \hfill
    \begin{subfigure}{0.49\textwidth}
        \centering
        \includegraphics[width=1.0\textwidth,]{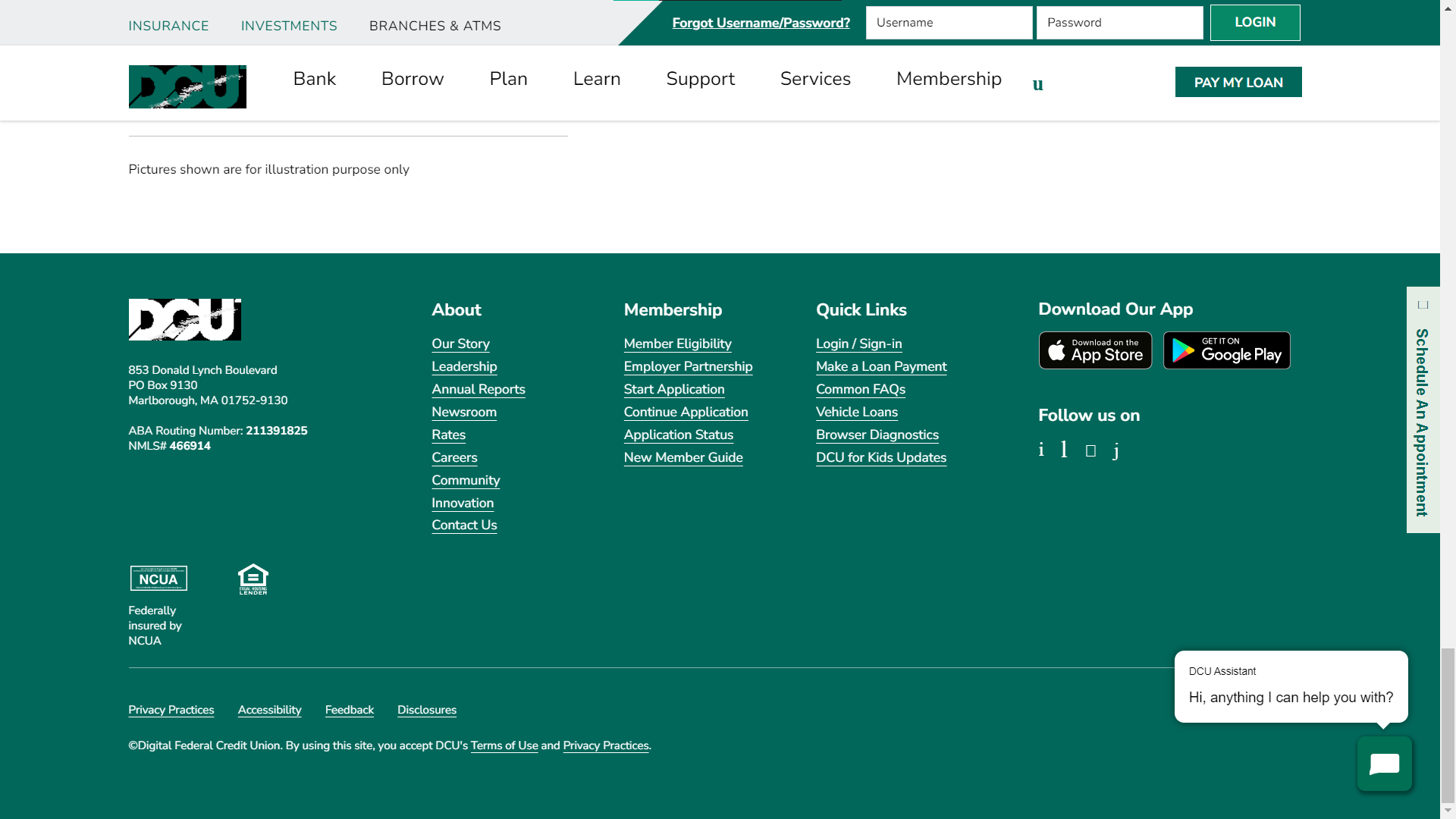}
        \caption{Generated Phishing webpage}
        \label{fig:generated_phishing_webpage_user_study}
    \end{subfigure}
    \caption{User Study: (\ref{fig:considered_legitimate_webpage_user_study}) Legitimate and (\ref{fig:generated_phishing_webpage_user_study}) \texttt{PhishOracle}-generated Webpage Screenshots}
    \label{fig:PhishOracle_WebPage_user_study}
    \Description{User Study: Legitimate and \texttt{PhishOracle}-generated webpage}
\end{figure*}

\textbf{Research Question}: The core research question (RQ) of the study is:
\textit{Can a user classify a webpage as `phishing' or `legitimate' when presented with a screenshot with its address bar masked?} To answer this, participants were shown webpage screenshots and asked to classify them as either phishing or legitimate. The goal is to assess whether users can distinguish phishing from legitimate webpages based solely on visual elements, without relying on the URL. While we acknowledge that real-world users typically have access to the address bar, prior report~\citep{ID_Agent_Report} has shown that many do not actively use URLs to verify a website’s legitimacy. This report found that $25.5 \%$ of users clicked on phishing links, and $18 \%$ went further to submit their credentials, indicating that a significant portion of users do not carefully inspect URLs before interacting with webpages.

\textbf{Setup}: We initiate the experiment by selecting $100$ legitimate webpages and generating $100$ corresponding adversarial phishing webpages using \texttt{PhishOracle}, by embedding randomly selected visual-based features (refer Table~\ref{tab:Phishing_Features}). Figure~\ref{fig:considered_legitimate_webpage_user_study} and Figure~\ref{fig:generated_phishing_webpage_user_study} showcase exemplar screenshots of legitimate and \texttt{PhishOracle}-generated phishing webpages respectively, employed in the user study. The phishing webpage screenshot (Figure~\ref{fig:generated_phishing_webpage_user_study}) is generated by embedding \textit{V4} visual-based phishing feature into the corresponding legitimate webpage (Figure~\ref{fig:considered_legitimate_webpage_user_study}).

\textbf{Distribution of Samples}: To facilitate this, we utilize Google Forms, offering two options for each webpage screenshot: ``Phishing'' or ``Legitimate''. We create $11$ Google Forms by randomizing the arrangement of legitimate and \texttt{PhishOracle}-generated adversarial phishing webpage screenshots. Employing Google Forms, we administer the user study to the $52$ participants. The first nine forms comprise a mix of $10$ legitimate and $15$ generated phishing webpage screenshots. One form featured $10$ legitimate and $16$ generated phishing webpage screenshots, while another included $10$ legitimate and $20$ generated phishing webpage screenshots. The URL field in each webpage screenshot is masked to ensure unbiased classification outcomes.

\begin{figure}[ht]
    \centering
    \includegraphics[width=1.0\textwidth, height=0.55\textwidth]{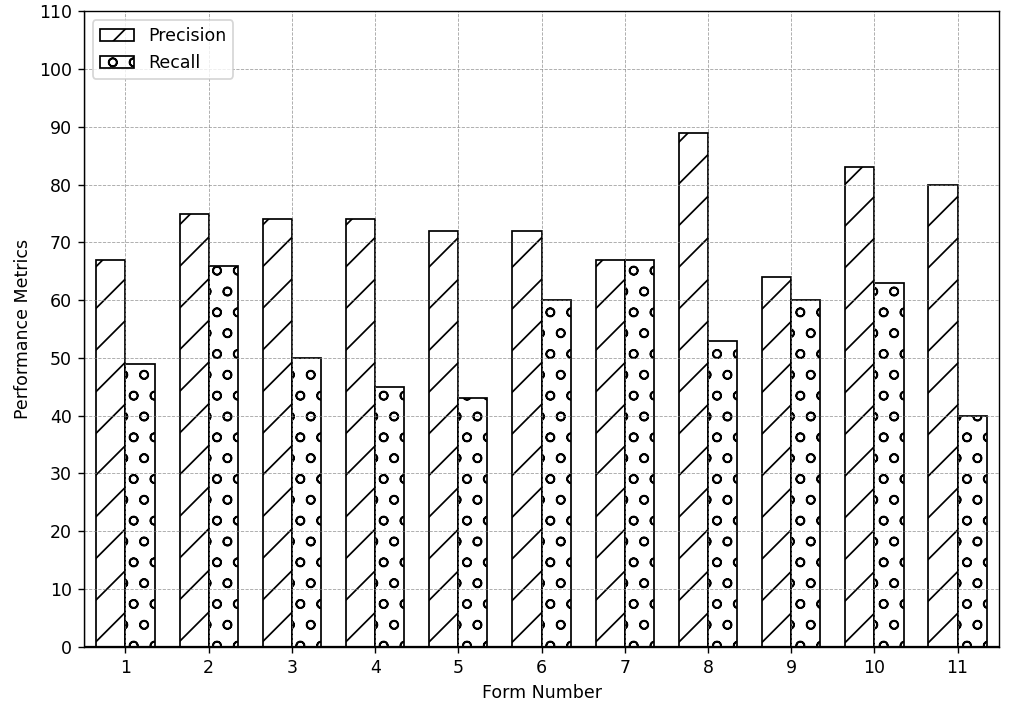}
    \caption{Results of our User Study}
    \label{fig:user_study}
    \Description{Results of our User Study}
\end{figure}

\textbf{Results}: Evaluating the feedback of participants reveals that, on average, $\sim$$48 \%$ of the phishing webpages generated by \texttt{PhishOracle} are misclassified as legitimate. The user study results are presented in Figure~\ref{fig:user_study}.

\textbf{Observations}: Our findings from the experiments (refer Section~\ref{sec:Validating_Phishing_Detection_Approaches}) and the user study emphasize that the logo transformation techniques (such as opacity, rotation, blur, noise, grey-scale mesh) not only affect the performance of the phishing webpage detection models but also deceive the users. This aligns with the work by Ying \textit{et al.}~\citep{yuan2024adversarial}, who argue that assessing the user's perception of adversarial phishing webpages is a necessary step in phishing webpage detection, as users are the actual targets.

\limitations{In our user study, the webpage screenshots do not include URLs, which could assist users in assessing the overall legitimacy of the webpage. Instead, users are presented with webpage screenshots and tasked to classify the webpage as `phishing' or `legitimate' based solely on visual appearance.}

\subsection{\texttt{PhishOracle} Web App}
\label{sec:PhishOracle_Web_App}
We next introduce our \texttt{PhishOracle} web app, an interactive tool developed to serve both user education and phishing research purposes. This web app facilitates the process of generating phishing webpages by utilizing a legitimate input URL.
\begin{figure*}[th]
    \centering
    \begin{subfigure}{0.24\textwidth}
        \centering
        \includegraphics[width=1.0\textwidth,]{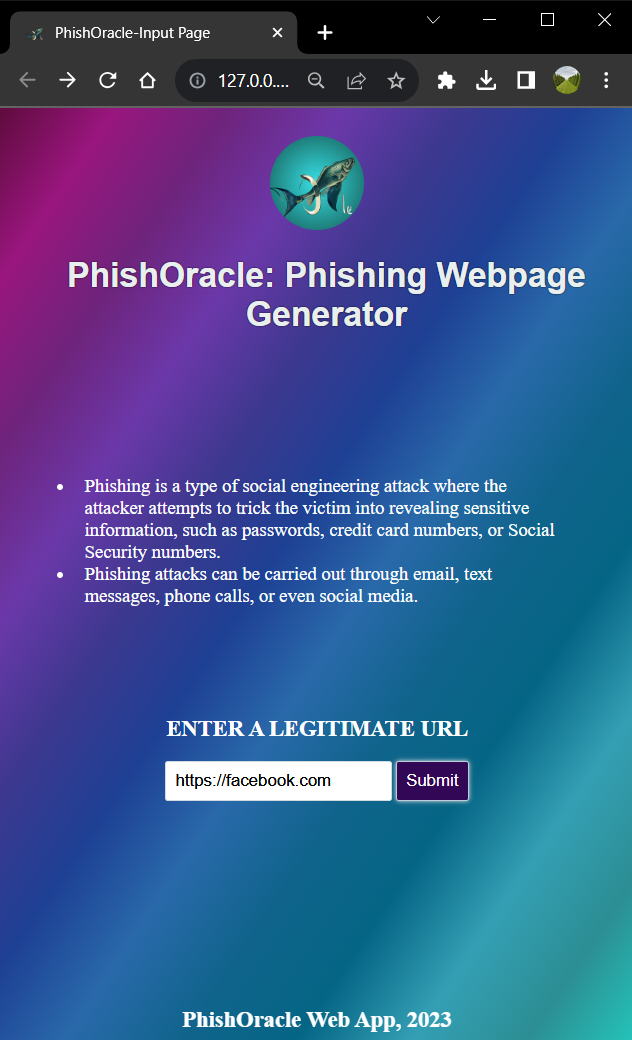}
        \caption{Index Page}
        \label{fig:PhishOracle_Web_App_Index_Page}
    \end{subfigure}
    \hfill
    \begin{subfigure}{0.24\textwidth}
        \centering
        \includegraphics[width=1.0\textwidth,]{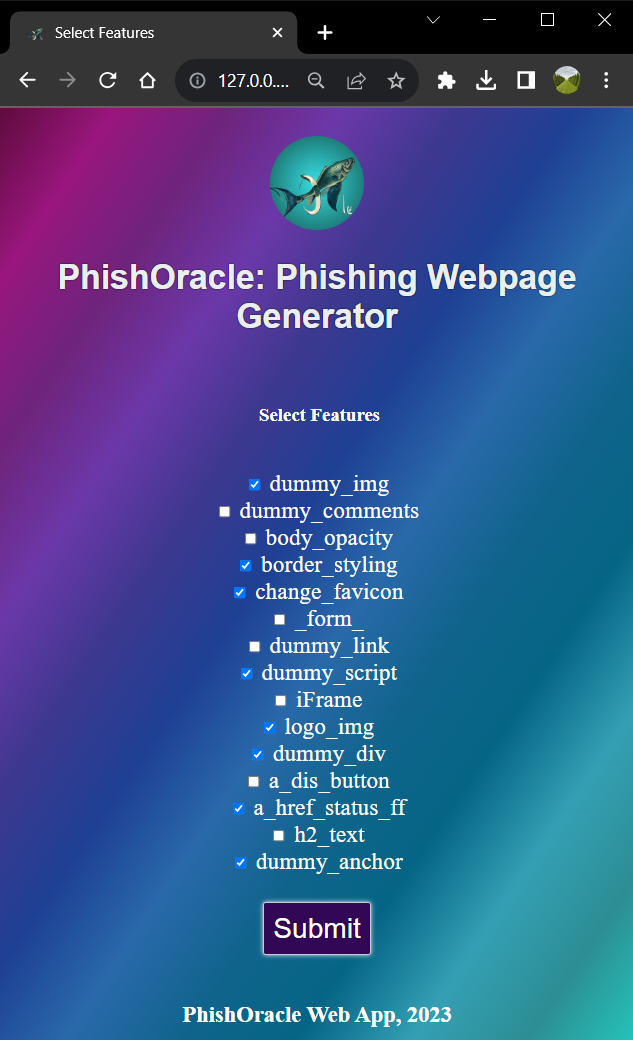}
        \caption{Feature Selection Page}
        \label{fig:PhishOracle_Web_App_Select_Features_Page}
    \end{subfigure}
    \begin{subfigure}{0.24\textwidth}
        \centering
        \includegraphics[width=1.0\textwidth,]{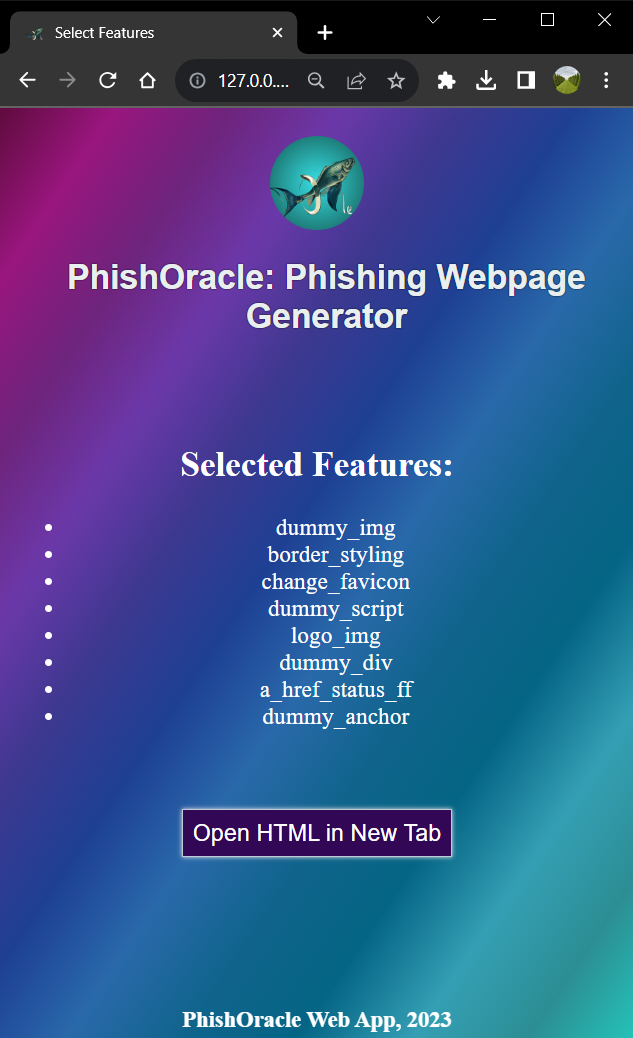}
        \caption{Generate Page}
        \label{fig:PhishOracle_Web_App_Generate_Page}
    \end{subfigure}
    \hfill
    \begin{subfigure}{0.24\textwidth}
        \centering
        \includegraphics[width=1.0\textwidth,]{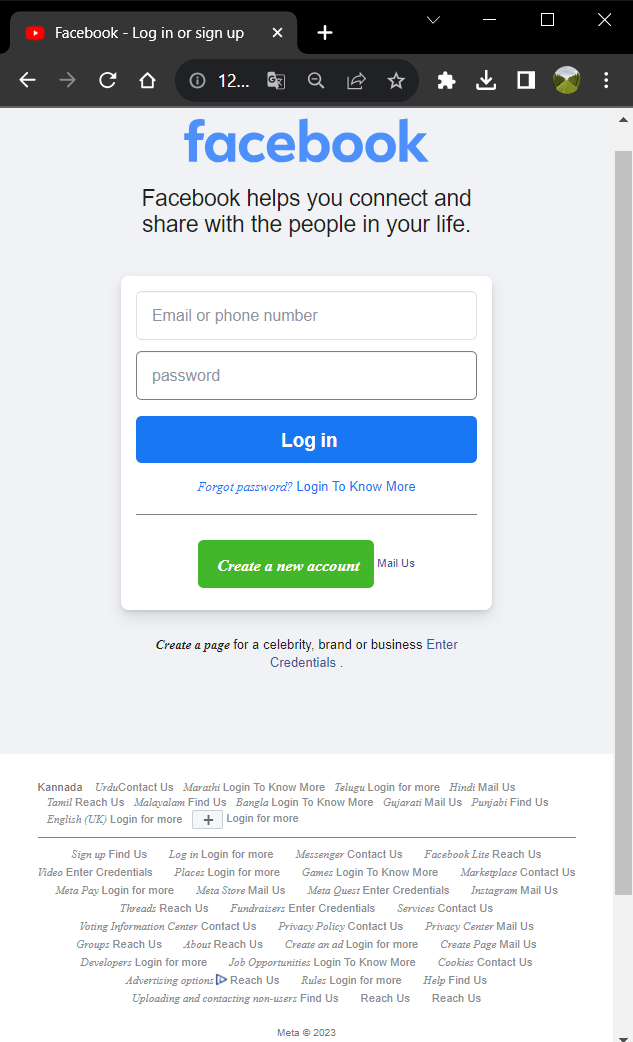}
        \caption{Phishing Webpage}
        \label{fig:PhishOracle_Web_App_Phishing_Page}
    \end{subfigure}
    \caption{Generating Phishing Webpage via \texttt{PhishOracle} Web App}
    \label{fig:PhishOracle_Web_App}
    \Description{Generating Phishing Webpage via \texttt{PhishOracle} Web App}
\end{figure*}

The web app was initially hosted on PythonAnywhere\footnote{\label{PythonAnywhere}PythonAnywhere, \url{https://www.pythonanywhere.com/}}, a cloud-based platform designed for hosting Python web apps, ensuring accessibility and reliability. However, the hosted web app was disabled due to a violation of the Terms and Conditions of PythonAnywhere as the content is related to phishing. We then hosted the web app on GitHub pages\footnote{\label{GitHubPages}GitHub Pages, \url{https://pages.github.com/}}, but the generated phishing webpage was marked ``Dangerous Site'' because of the application containing phishing activities. The web app is thus added as a repository on GitHub~\citep{PHISHORACLE_WEBAPP_GITHUB} which can be downloaded and hosted on a local system. The web app comprises three rendering HTML templates: \texttt{index.html}, \texttt{select\_features.html}, and \texttt{generate\_webpage.html}. The \texttt{index.html} (Figure~\ref{fig:PhishOracle_Web_App_Index_Page}) serves as the user's entry point, offering an input text box for legitimate URL submission. Once a URL is provided, the application fetches the source code of the associated webpage, extracts relevant features, and presents them as checkboxes on the \texttt{select\_features.html} page (Figure~\ref{fig:PhishOracle_Web_App_Select_Features_Page}), allowing users to select features to incorporate into the webpage. After selecting features, the application embeds the chosen features into the legitimate webpage to generate the corresponding phishing webpage, which can be viewed on the \texttt{generate\_webpage.html} page (Figure~\ref{fig:PhishOracle_Web_App_Generate_Page}). Users can open the generated phishing webpage (Figure~\ref{fig:PhishOracle_Web_App_Phishing_Page}) in a new browser tab with a single click. \texttt{PhishOracle} web app streamlines the process of generating phishing webpages for research and testing, offering a user-friendly interface for feature selection and webpage generation.

\section{Limitations}
\label{sec:Limitations}
While our black-box approach demonstrates the effectiveness of visually transformed adversarial phishing webpages in evading detection models, it comes with some limitations. Our work focuses on a limited subset of transformation techniques and does not consider structural modifications such as background or layout changes. Although on the positive side, the selected transformations are simple to implement, visually deceptive to users, and less complex for real-world adversaries with limited knowledge. Additionally, we evaluated only one LLM-based phishing detection system. Evaluating against a broader set of LLMs is needed to better understand the generalizability of our approach to modern LLM-based phishing webpage detections.

\section{Conclusion}
\label{sec:Conclusion_and_Future_Scope}
In this paper, we developed \texttt{PhishOracle} to generate adversarial phishing webpages by embedding randomly selected content-based and visual-based phishing features into legitimate webpages. These adversarial phishing webpages were used to evaluate the robustness of the Stack model, VisualPhishNet, Phishpedia, and an LLM-based phishing detector. Our findings reveal that the Stack model exhibits reduced performance, while both VisualPhishNet and Phishpedia incorrectly identify the brands on adversarial phishing webpages due to various logo transformation techniques, leading to incorrect classifications. On the other hand, LLM-based phishing detection using Gemini (\textit{LLM-PD\textsuperscript{S}}) demonstrates better performance against these adversarial attacks, although its detection capability also drops due to \texttt{PhishOracle}-generated phishing pages. These results highlight the vulnerabilities of existing phishing webpage detection models, emphasizing the potential of MLLMs in advancing phishing webpage detection. Furthermore, our evaluation of tools from different security vendors on VirusTotal shows that adversarial phishing webpages generated by \texttt{PhishOracle} remain undetected within the first $24$ hours, leaving them accessible to potential victims for an extended period. Moreover, the user study demonstrates that on an average $\sim$$48 \%$ of the \texttt{PhishOracle}-generated adversarial phishing webpages are misclassified as legitimate, by the participants. The datasets, \texttt{PhishOracle} code and web app are available on our GitHub repositories~\citep{PHISHORACLE_GITHUB, PHISHORACLE_WEBAPP_GITHUB}. \texttt{PhishOracle} web app empowers users to clone and host it locally, enabling them to input a legitimate URL, choose specific phishing features, and generate corresponding phishing webpages.

The visual-based features in our tool add opacity to the logos and utilize transformation techniques such as rotation, blurring, adding grey-colored mesh, and noise on logos within a webpage. In our future work, we plan to add a few more techniques to generate adversarial logos and incorporate them as a feature in \texttt{PhishOracle} tool.

\textbf{Ethical Statement.} Our institution does not require any formal IRB approval to carry out the research discussed here. Adhering to the principles outlined in the Menlo report~\citep{bailey2012menlo}, our user study is conducted with strict adherence to ethical guidelines, ensuring the non-forging and intentional avoidance of collecting sensitive information from participants. The study involved participants selecting an option (`Phishing' or `Legitimate') based on the visual appearance of webpage screenshots. Although we plan to release the source code of \texttt{PhishOracle} tool on GitHub repositories~\citep{PHISHORACLE_GITHUB, PHISHORACLE_WEBAPP_GITHUB}, rather than making it publicly available without restrictions, we will provide access on `Request Access' basis to ensure responsible use. It is intended solely for educational and research purposes and not for any illegal or unethical activities.


\bibliographystyle{unsrt}
\balance
\bibliography{references}










    \end{document}